\documentclass[11pt]{article} 
  
\usepackage{amsmath,amssymb}
\usepackage{tikz}
\usepackage{pgfplots}
\usetikzlibrary{decorations.pathreplacing}
\usepackage{multicol}
\usepackage{nicefrac}

\newboolean{generateTikzPics} 
\setboolean{generateTikzPics}{false} 
\newcommand{\inputTikz}[2]
{
\ifthenelse{\boolean{generateTikzPics}}{
 \tikzsetnextfilename{#2}
 \input{#1}
 }
{
 \includegraphics[scale=1]{externalized/#2}
}
}
\newcommand{\ownCBar}[6]
{
\ifthenelse{\boolean{generateTikzPics}}{
 \tikzsetnextfilename{#5}
\begin{tikzpicture}
\pgfplotscolorbardrawstandalone[
    colormap/jet,
    colorbar horizontal,
    point meta min=#3,
    point meta max=#4,
    scaled ticks=false,
    colorbar style={ xtick={#6}},
    x tick label style={
        /pgf/number format/.cd,
            fixed,
            precision=3,
        /tikz/.cd
    },
    colorbar style={
        width=0.5\textwidth,
	}]
\end{tikzpicture}
}
{
\includegraphics{externalized/#5}
}
}

\usepackage{pgfplots, pgfplotstable}
\pgfplotsset{compat=newest}
\usepgfplotslibrary{patchplots}
\usetikzlibrary{shadings}
\usetikzlibrary{shapes}
\usetikzlibrary{positioning,arrows}
\usepgfplotslibrary{fillbetween}
\usetikzlibrary{backgrounds}
\usepackage{tikz}
\usepackage{tikz-3dplot}
\usetikzlibrary{calc}
\usetikzlibrary{3d}
\usetikzlibrary{external}
\usepackage{siunitx}  
\usepackage{subcaption}

\usepackage[normalem]{ulem}

\newcounter{tony}
\newcommand{\be}{\mbox{\boldmath{$e$}}}
\newcommand{\fb}{\mbox{\boldmath{$f$}}}

\newcommand{\bm}{\mbox{\boldmath{$m$}}}
\newcommand{\bn}{\mbox{\boldmath{$n$}}}

\newcommand{\bp}{\mbox{\boldmath{$p$}}}

\newcommand{\bs}{\mbox{\boldmath{$s$}}}
\newcommand{\bt}{\mbox{\boldmath{$t$}}}
\newcommand{\bu}{\mbox{\boldmath{$u$}}}
\newcommand{\bv}{\mbox{\boldmath{$v$}}}

\newcommand{\bC}{\mbox{\boldmath{$C$}}}

\newcommand{\bU}{\mbox{\boldmath{$U$}}}

\newcommand{\BBR}{\mbox{$\mathbb{R}$}}

\newcommand{\bzero}{\mbox{$\bf 0$}}

\newcommand{\bepsilon}{\mbox{\boldmath{$\varepsilon$}}}

\newcommand{\bsigma}{\mbox{\boldmath{$\sigma$}}}

\newcommand{\half}{\mbox{$\frac{1}{2}$}}

\newcommand{\equref}[1]{(\ref{eq:#1})}

\newcommand{\beq}{\begin{equation}}

\newcommand{\beqna}{\begin{eqnarray}}
\newcommand{\eeqna}{\end{eqnarray}}

\begin{document} 
\begin{center}
{\large {\bf A thermodynamically consistent theory of stress-gradient plasticity}}\vspace{3ex}\\
B D Reddy$^{1}$, P Steinmann$^{2,3}$ and A Kerga{\ss}ner$^{2}$ \\
$^1$ Centre for Research in Computational and Applied Mechanics, University of Cape Town, 7701 Rondebosch, South Africa\\
$^2$Institute of Applied Mechanics, Friedrich-Alexander Universit\"{a}t Erlangen-N\"{u}rnberg, 91054 Erlangen, Germany\\
$^3$ Glasgow Computational Engineering Centre, University of Glasgow, G12 8QQ Glasgow, United Kingdom
\vspace{6ex}\\ 
{\em The authors dedicate this work to the memory of Hussein M Zbib}
\vspace{6ex}\\ 
\today
\end{center}
   
{\bf Abstract}

As an extension to strain-gradient models of size-dependent plastic behaviour, this work proposes a model for a stress-gradient theory. The model is distinguished from earlier works on the topic by its being embedded in a thermodynamically consistent framework. The development is carried out in the context of single-crystal plasticity, and draws on thermodynamically consistent models for single-crystal conventional and strain-gradient plasticity. The model is explored numerically using the example of torsion of a thin wire comprising a face centred cubic crystal, and its behaviour compared with that based on a recent disequilibrium density model of size-dependent plasticity. 

\section{Introduction}
There is abundant experimental evidence of the size-dependent response to mechanical loading of crystalline materials at the micro-scale. A non-exhaustive list includes the works \cite{fleck1994mah,stoelken1998e,uchic2004dfn,ehrler2008,dunstan2009,liu2012,liu2013}. Size-dependence manifests typically as both strengthening, that is, an increase in incipient yield stress, and hardening, that is, an increase in the slope of the load-deformation relation in the plastic range, in both cases with respect to decrease in sample size. These effects are attributed to the presence of geometrically necessary dislocations (GNDs) in non-homogeneous deformations: these form obstacles to further unhindered flow of dislocations that represent the key mechanism of plasticity.  Consequently, trapping of dislocations results in increased hardening \cite{fleck1994mah}.

The continuum concept of dislocation densities originates in the early works \cite{nye1953,bilby1955bs,kroner1958,kroner1959,kroner1959s} (see the work \cite{steinmann2015} for an account on continuum defect densities in differential geometry).  In local continuum plasticity, hardening due to statistically stored dislocations (SSDs) is routinely modelled in terms of (local) internal variables, so that the consideration of SSDs in continuum plasticity is unable to capture size-dependence. In formulations of continuum plasticity, hardening due to GNDs, on the other hand, requires the consideration of the continuum density of GNDs. This feature is often captured in continuum models by the inclusion of a dependence on plastic strain gradient that introduces a length scale into the formulation. A sample of such approaches includes the works
\cite{
zbib1992a,steinmann1996,fleck1997h,menzel2000s,
gurtin2002,
gudmundson2004,
gurtin2007al,
evans2009h,forest2009,forest2010a,reddy20111,reddy20112,gurtin2014r,mcbride2018rs}, among many others.

Strain-gradient theories fall typically into two models, which may appear singly or in combined form: energetic theories, in which plastic strain gradients are captured through inclusion of a recoverable energy term; and dissipative models, in which plastic strain gradients are captured directly in a flow relation. Dissipative models give rise to size effects in incipient yield and subsequent hardening rate. Energetic models in which the defect energy depends on the accumulated plastic strain or an equivalent accumulated measure display a similar strengthening response \cite{Ohno-Okumura2007,Ohno-Okumura-Shibata2008}. This relationship is evident from the equivalence between dissipative models and energetic models in which the defect energy depends on the accumulated plastic strain or dislocation density (see for example \cite{Gurtin-Reddy2009} for the polycrystalline case and \cite{reddy20112} for the case of single-crystal gradient plasticity). On the other hand, 
energetic models with a polynomial, for example quadratic defect energy, introduce a back-stress in the yield condition and lead to size effects only in the rate of hardening (see for example \cite{Idiart-Fleck2010,chiricotto2012gt,chiricotto2016gt}).
In simple spring-dashpot models it has been shown \cite{mcbride2018rs} that the behaviour of the energetic model may be captured in a simple analogue context by a spring and slider in series, while for the dissipative model the analogous arrangement would be that of a spring and slider in parallel.



Models of an alternative approach, which relates size-dependent responses to gradients in stress, have been proposed and explored in recent years, an early approach being that due to Hirth \cite{hirth2006}. In the important contribution by Chakravarty and Curtin \cite{chakravarthy-curtin2011}, a mechanistically motivated model of stress-gradient plasticity is developed. The model is based on the strengthening that arises when a stress gradient acts over configurations comprising dislocation sources and obstacles. The resultant length scale has a clear physical interpretation, that is, average obstacle spacing. The key features are incorporated into a continuum model, and a range of numerical results using this model compared against those obtained in experiments on torsion of thin wires and bending of thin films, with good to excellent agreement observed. In particular, both size-dependent strengthening and hardening are observed, and captured by the model. The work \cite{liu2014hzs} develops a stress-gradient theory that accounts for both grain and specimen sizes in determination of the length scale. Results obtained from the continuum theory show good agreement with experimental results from torsion and bending tests.

While the model presented in \cite{chakravarthy-curtin2011} has a firm physical basis, and leads to good agreement with experimental results, it lacks a framework of thermodynamic consistency, as acknowledged by the authors. The objective of this work is to propose a model for a stress-gradient theory, motivated by that in \cite{chakravarthy-curtin2011}, and embedded in a thermodynamically consistent framework. The development is carried out in the context of single-crystal plasticity, and draws on a thermodynamically consistent model for single-crystal conventional and strain-gradient plasticity \cite{reddy20112}. Models of strain-gradient plasticity such as that proposed in \cite{gurtin2002} make provision for microstresses through a principal virtual power relation, and a resultant microforce balance equation. Such additional quantities and associated equations are not present in the stress-gradient theory; nor, as pointed out in \cite{chakravarthy-curtin2011}, is it necessary to impose higher-order boundary conditions on the plastic strain gradient.

The structure of the rest of this work is as follows. The framework for rate-independent single-crystal elastoplasticity is presented in Section 2. In particular, the flow relation is based on the notion of an associative flow law or, equivalently, a positively homogeneous dissipation function. The extension to viscoplasticity is recorded, as the computational work makes use of a viscoplastic approximation. The extension to stress-gradient plasticity is presented in Section 3, starting with the adoption of a yield function that includes stress gradients, and proceeding to extend the conventional model. The stress gradients appearing in the yield function are treated as quantities conjugate to an internal variable, and it is shown that these may be obtained through a weak or variational formulation involving the stress. The weak formulation makes it possible to approximate only the stress, and not its gradient, in finite element approximations of the problem. Section 4 presents the results using the new model, in the example of torsion of a thin wire. Though the precise values of incipient yield stress are difficult to determine, there is clear size-dependent strengthening behaviour, with mild hardening. The work concludes with summarizing remarks and an indication of future work in Section 5.
                   
\section{The classical case}

\subsection{The rate-independent problem}\label{rate-ind}
In this section we give an overview of a thermodynamically consistent model for classical single-crystal plasticity with isotropic hardening, as presented, for example, in \cite{Han-Reddy2013}. The corresponding formulation for stress-gradient plasticity will follow this framework closely.

Within a small-strain context the total and plastic strains are denoted respectively by $\bepsilon$ and $\bp$, and the latter is given by
\begin{equation}
\bp = \mbox{$\frac{1}{2}$}\sum_A \gamma^\alpha [\bs^\alpha \otimes \bm^\alpha + \bm^\alpha \otimes \bs^\alpha]\,,
\label{p}
\end{equation}
where $\gamma^\alpha$ is the slip on slip plane $\alpha$, and $\bs^\alpha$ and $\bm^\alpha$ are respectively the slip direction and slip plane normal on this plane.

The free energy $\psi$ is a function of the elastic strain $\be = \bepsilon -\bp$ and of a set of scalar hardening variables $\eta^\alpha\ (\alpha = 1,\ldots , A)$, 
one for each of the slip systems. The separable form
\begin{equation}
\psi = \psi^e (\be) + \psi^h (\vec{\eta})
\label{eq:psisingle}
\end{equation}
is assumed. Here, and henceforth, $\vec{\eta}$ denotes the array $(\eta^1,\ldots , \eta^A)$.
The elastic relation is given by 
\begin{equation}
\bsigma = {\bC}[\bepsilon - \bp] = {\bC}\be = \frac{\partial \psi^e}{\partial \be}
\label{eq:elastreln}
\end{equation}
where $\bsigma$ denotes the Cauchy stress. We assume cubic symmetry, so that the elasticity tensor $\bC$ has components  
\begin{equation}
\bC = 
\begin{bmatrix}
C_{11} & C_{12} & C_{12} &&&\\
C_{12} & C_{11} & C_{12} &&&\\
C_{12} & C_{12} & C_{11} &&&\\
&&&C_{44} &&\\
&&&&C_{44} &\\
&&&&&C_{44} \\
\end{bmatrix}
\label{elastcubic}
\end{equation}
where
\begin{equation}
C_{11} = E\frac{1-\nu}{[1+\nu][1+2\nu]}, \quad
C_{12} = E\frac{\nu}{[1+\nu][1+2\nu]}, \quad
C_{44} = \mu\, .
\label{eq:coeffCubSym}
\end{equation}
where $E$ is Young's modulus, $\nu$ is Poisson's ratio, and $\mu$ is the shear modulus. 

We use the reduced dissipation inequality 
\begin{equation}
\dot{\psi} - \bsigma:\dot\bepsilon \leq 0
\label{eq:dissineq}
\end{equation}
together with the elastic relation and the form \eqref{eq:psisingle} for $\psi$, to obtain
\begin{equation}
-\bsigma:\dot{\bp} + \sum_\alpha \frac{\partial \psi^h}{\partial\eta^\alpha}\dot{\eta}^\alpha \leq 0\,.
\label{eq:3.83}
\end{equation}
The first term on the left-hand side of this inequality is, from \eqref{p} and the symmetry of $\bsigma$,
\begin{equation}
\bsigma:\dot{\bp}  = \bsigma :\sum_\alpha \dot{\gamma}^\alpha\bs^\alpha\otimes \bm^\alpha 
 = \sum_\alpha \dot{\gamma}^\alpha\bsigma:[\bs^\alpha\otimes \bm^\alpha ]
 = \sum_\alpha \dot{\gamma}^\alpha\tau^\alpha\,,
\label{eq:plasticwork}
\end{equation}
where
\begin{equation}
\tau^\alpha := \bsigma :[\bs^\alpha \otimes \bm^\alpha ] = \bsigma\bm^\alpha\cdot\bs^\alpha
\label{eq:resshear1}
\end{equation}
is the {\em resolved shear stress}\index{resolved shear stress} on the plane with normal $\bm^\alpha$.
Furthermore, the thermodynamic variable $g^\alpha$ conjugate to 
$\eta^\alpha$ is defined by 
\begin{equation}
g^\alpha := - \frac{\partial \psi^h}{\partial \eta^\alpha}\,, 
\label{eq:galpha}
\end{equation}
and we further stipulate that
\begin{equation}
g^\alpha \leq 0\,.
\label{gleq0}
\end{equation}
It follows that the reduced dissipation inequality \eqref{eq:3.83} becomes
\begin{equation}
\sum_\alpha \left[\tau^\alpha\dot{\gamma}^\alpha+g^\alpha\dot{\eta}^\alpha\right]\geq 0\,.
\label{eq:dissineq1}
\end{equation} 

For example, for the  case of a quadratic hardening term, that is,
\begin{equation}
\psi^h(\vec{\eta}) = \half k \sum_\alpha [\eta^\alpha]^2\,,
\label{eq:psihquadratic}
\end{equation}
the thermodynamic conjugate, which has the dimensions of stress, is given by
\begin{equation}
g^\alpha = -k\eta^\alpha\,.
\label{eq:g^alpha_lin}
\end{equation}

\noindent {\bf Yield.}
We define on the $\alpha$th slip system the yield condition
\begin{equation}
\phi (\tau^\alpha,g^\alpha) = |\tau^\alpha | - [\tau_0 - g^\alpha ] \leq 0\,.
\label{eq:yieldalpha}
\end{equation}
Here $\tau_0$ is an initial yield stress, assumed for convenience to be the same for 
all slip systems. 
The elastic region ${\cal E}$ 
is now defined to be the intersection of elastic regions for all the slip systems: 
\begin{equation}
{\cal E} = \bigcap_\alpha {\cal E}^\alpha 
=\bigcap_\alpha \left\{(\bsigma,\vec{g}):\ \phi(\tau^\alpha,g^\alpha )< 0\,,\ \alpha = 1,\ldots,A\right\}\,.
\label{eq:E}
\end{equation}
The elastic region defined by \equref{E} is clearly convex. Next, the assumption of a normality law gives
\begin{subequations}
\begin{align}
\dot{\gamma}^\alpha & = \lambda^\alpha \frac{\partial \phi}{\partial \tau^\alpha}\,,  
\label{eq:normalitySCa}\\
\dot{\eta}^\alpha & = \lambda^\alpha \frac{\partial \phi}{\partial g^\alpha} = \lambda^\alpha
\label{eq:normalitySCb}
\end{align}
\label{eq:normality}
\end{subequations}
together with the complementarity relations
\begin{equation}
\lambda^\alpha \geq 0\,,\qquad \phi (\tau^\alpha,g^\alpha)  \leq 0\,,
\qquad \lambda^\alpha\phi  (\tau^\alpha,g^\alpha)  = 0\,.
\label{eq:comp}
\end{equation}
From (\ref{eq:normalitySCa}) and (\ref{eq:normalitySCb}) it follows that
\begin{equation}
\dot{\eta}^\alpha = |\dot{\gamma}^\alpha| = \lambda^\alpha\,.
\label{eq:mudotmodgammadot}
\end{equation}

{\bf The flow relation in terms of the dissipation function.}\ \ The dual of the flow relation \equref{normality} is given by
\begin{equation}
(\tau^\alpha,g^\alpha) \in \partial D(\dot{\gamma}^\alpha,\dot{\eta^\alpha}), 
\label{flow_ito_D}
\end{equation}
where $\partial D$ denotes the subdifferential of $D$, defined to be the sets of pairs $(\tau^\alpha,g^\alpha)$ that satisfy   
\begin{equation}
D(\tilde{\gamma}^\alpha,\tilde{\eta}^\alpha) \geq D(\dot{\gamma}^\alpha,\dot{\eta}^\alpha) + \tau^\alpha [\tilde{\gamma}^\alpha - \dot{\gamma}^\alpha]
+ g^\alpha [\tilde{\eta}^\alpha - \dot{\eta}^\alpha]\qquad \mbox{for all} \ (\tilde{\gamma}^\alpha,\tilde{\eta}^\alpha)\,.
\label{flow_ito_Dfull}
\end{equation}

The dissipation function $D(\tilde{\gamma}^\alpha,\tilde{\eta}^\alpha)$ is given by (see \cite{Han-Reddy2013} for a derivation)
\begin{equation}
D(\tilde{\gamma}^\alpha,\tilde{\eta}^\alpha) = \left\{ \begin{array}{ll}
\tau_0 |\tilde{\gamma}^\alpha| & \quad \mbox{if}\ |\tilde{\gamma}^\alpha| \leq \tilde \eta^\alpha\,,\\
+ \infty &  \quad \mbox{if}\ |\tilde{\gamma}^\alpha| > \tilde \eta^\alpha\,.
\end{array} \right.
\label{dissfn}
\end{equation}

{\bf The effective dissipation function.}\ \ Restricting attention to the subset of arbitrary slips and internal variables that satisfy
\[ 
|\tilde{\gamma}^\alpha| = \tilde{\eta}^\alpha, 
\]
rearrangement of the inequality \eqref{flow_ito_Dfull} and the use of \eqref{dissfn}$_1$ gives
\begin{eqnarray}
\tau_0 |\tilde{\gamma}^\alpha | - g^\alpha\tilde{\eta}^\alpha \geq \tau_0 |\dot{\gamma}^\alpha| - g^\alpha\dot{\eta}^\alpha+ \tau^\alpha [\tilde{\gamma}^\alpha - \dot{\gamma}^\alpha]
\label{Dform1}
\end{eqnarray}
or, given the constraints on the admissible generalized quantities,
\begin{eqnarray}
\left[\tau_0-g^\alpha\right]|\tilde{\gamma}^\alpha|\geq\left[\tau_0-g^\alpha\right]|\dot{\gamma}^\alpha|+ \tau^\alpha [\tilde{\gamma}^\alpha - \dot{\gamma}^\alpha]\,.
\label{Dform2}
\end{eqnarray}
Here we have used the fact that $\dot{\eta}^\alpha=|\dot{\gamma}^\alpha|$. 
We define the {\em effective} dissipation function 
\begin{equation}
D_{\rm eff}(\tilde{\gamma}^\alpha,g^\alpha) := \left[\tau_0 - g^\alpha\right] |\tilde{\gamma}^\alpha|\,.
\label{Dred1}
\end{equation}
Then the flow law may be expressed in terms of a dissipation function which depends only 
on the plastic slips, with the conjugate quantity $g^\alpha$ being treated implicitly as a 
function of $\eta^\alpha$. That is, we can write, from \eqref{Dform2} and \eqref{Dred1}, 
\begin{equation}
\tau^\alpha \in \partial_1 D_{\rm eff} (\dot{\gamma}^\alpha,g^\alpha) 
\label{subdiff}
\end{equation}
where $\partial_1 D_{\rm eff}$ denotes the subdifferential of $D_{\rm eff}$ 
with respect to its first argument; that is, 
\begin{equation}
D_{\rm eff}(\tilde{\gamma}^\alpha,\tilde{g}^\alpha) \geq D_{\rm eff} (\dot{\gamma}^\alpha,g^\alpha) 
+ \tau^\alpha [\tilde{\gamma}^\alpha - \dot{\gamma}^\alpha ]\,.
\label{eq:flowDeffSC}
\end{equation}
In particular, when plastic flow occurs on slip system $\alpha$ so that 
$\dot{\gamma}^\alpha \neq 0$, then
\begin{equation}
\tau^\alpha = \left.\frac{\partial D_{\rm eff}(\tilde\gamma^\alpha,g^\alpha)}{\partial \tilde\gamma^\alpha}
\right|_{\dot{\gamma}^\alpha}  = [\tau_0 - g^\alpha]\, \mbox{sgn}\,\dot{\gamma}^\alpha\,.
\label{eq:flowDeffSC2}
\end{equation}
This relation could be obtained directly from the flow relation \eqref{eq:normality}, as follows: we have
\begin{align}
\dot{\gamma}^\alpha & = |\dot{\gamma}^\alpha| \frac{\partial \phi}{\partial \tau^\alpha} \nonumber \\
& = |\dot{\gamma}^\alpha| \frac{\tau^\alpha}{|\tau^\alpha|} \nonumber \\
& = |\dot{\gamma}^\alpha| \frac{\tau^\alpha}{\tau_0 - g^\alpha}\,.
\end{align}
Rearrange to get
\begin{equation}
\tau^\alpha = \left. \frac{\partial D_{\rm eff}(\tilde{\gamma}^\alpha, g^\alpha)}{\partial \tilde{\gamma}^\alpha}\right|_{\tilde{\gamma}^\alpha = \dot{\gamma}^\alpha}\,.
\label{flow_eff}
\end{equation}
The same applies if $D_{\rm eff}$ is replaced by a smooth, for example viscoplastic approximation, in which case the inequality becomes an equation.

{\bf A more general form of hardening.} \ \ We can build into the flow relation a distinction between 
self-hardening, which characterizes hardening on a slip plane due to slip on all slip systems coplanar 
to the given plane; and latent hardening, which refers to hardening on a slip plane due to slip on all other individual slip planes. These are most easily captured in hardening relations by making use of the coplanarity moduli $\chi^{\alpha\beta}$ \cite{Gurtin-Fried-Anand2010}, defined by
\begin{equation}
\chi^{\alpha\beta} = \left\{ \begin{array}{cl} 1 & \mbox{for $\alpha$ and $\beta$ coplanar}\,; \\ 
0 & \mbox{otherwise}\,.
\end{array} \right.
\label{eq:coplanar}
\end{equation}
The slip systems $\alpha$ and $\beta$ are said to be coplanar if their corresponding
slip planes coincide.
Then the yield function can be extended to include self- and latent hardening by defining 
\begin{equation}
g^{\alpha}_\chi = -\sum_\beta |g^{\alpha\beta}| \big[\chi^{\alpha\beta} + q [1 - \chi^{\alpha\beta}]\big]\,,
\label{gchi}
\end{equation}
where $q$ is an interaction coefficient.

\subsection{Viscoplastic approximation}
We make use of a Norton-Hoff approximation of the yield condition.  Writing the yield condition \eqref{eq:yieldalpha} as a gauge, that is, in the form
\begin{equation}
\Phi(\tau^\alpha,g^\alpha ) := \frac{|\tau^\alpha|}{\tau_0 - g^\alpha} \leq 1\,,
\label{eq:Phi}
\end{equation}
the Norton-Hoff viscoplastic regularization of the flow law (see \cite{Friaa1978,Hoff1962,Norton1929}) amounts to replacing (\ref{eq:Phi}) by
\begin{align}
\Phi_p(\tau^\alpha,g^\alpha ) & = \frac{1}{p}[\Phi (\tau^\alpha,g^\alpha )]^p \nonumber \\
& = \frac{1}{p}\Big[\frac{|\tau^\alpha|}{\tau_0 - g^\alpha}\Big]^p
\label{eq:Phi_p}
\end{align}
where $p > 1$. Then the corresponding flow relation is
\begin{align}
\dot{\gamma}^\alpha & = \frac{\partial \Phi_p}{\partial \tau^\alpha} \nonumber \\
& = \frac{|\tau^\alpha|^{p-1}}{[\tau_0 - g^\alpha]^{p}}\mbox{sgn}\,\tau^\alpha\,.
\label{eq:phieta}
\end{align}
This relation may be inverted either directly, or by using the approximation of the dissipation function corresponding to $\Phi_p$. This is given by (see for example \cite{Ebobisse-Reddy2004} for a detailed treatment of such dual relations)
\begin{align}
D_{{\rm eff},p'}(\tilde{\gamma}^\alpha) = \frac{1}{p'} [\tau_0 - g^\alpha]|\tilde{\gamma}^\alpha|^{p'}\,,
\label{Deffq}
\end{align}
where $1/p' = 1 - 1/p$. Then the stress is given by 
\begin{align}
\tau^\alpha & = \left.\frac{\partial D_{{\rm eff},p'}}{\partial \tilde{\gamma}^\alpha}\right|_{\dot{\gamma}^\alpha} \nonumber \\
& = [\tau_0 - g^\alpha]|\dot{\gamma}^\alpha|^{p'-1}\mbox{sgn}\,\dot{\gamma}^\alpha\,.
\label{Flowq}
\end{align}

%

\subsection{The equilibrium equation and boundary conditions}\label{equil+bcs}

To complete the formulation of the problem we add the weak form of the equilibrium equation. The body occupies a domain $\Omega \subset \BBR^d\ (d = 2,3)$, with boundary $\partial \Omega$. Assume that the boundary conditions are
\begin{equation}
\bu = \bar{\bu}\quad\mbox{on}\ \partial\Omega_D,\qquad \bt = \bsigma\bn = \bar{\bt}\quad\mbox{on}\ \partial\Omega_N\,,
\label{macrobcs}
\end{equation}
where $\bu$ denotes the displacement, $\bt$ the surface traction, and $\bar{\bu}$ and $\bar{\bt}$ are respectively a prescribed displacement and traction on complementary parts $\partial \Omega_D$ and $\partial \Omega_N$ of $\partial \Omega$. Set $V: = \{ \bv\ |\ v_i \in H^1(\Omega),\ \bv = \bzero\ \mbox{on}\ \partial \Omega_D\}$, where $H^1(\Omega)$ is the Hilbert space of functions which together with their first weak derivatives are square-integrable. Define the function $\bar{\bU}$ such that $\bar{\bU} = \bar{\bu}$ on $\partial \Omega_D$. Then the equilibrium problem is as follows: find $\bu$ such that $\bu - \bar{\bU} \in V$ and 
\begin{equation}
\int_\Omega \bsigma (\bu,\vec{\gamma}):\bepsilon(\bv)\ dx = \int_{\partial \Omega_N}\bar{\bt}\cdot \bv\ ds +   \int_\Omega \fb\cdot\bv\ dx\qquad \mbox{for all}\ \ \bv \in V\,.
\label{weak_equil}
\end{equation}
Here $\fb$ is the prescribed body force on $\Omega$ and traction on $\partial \Omega_N$. 

The solution of \eqref{weak_equil} and \eqref{eq:phieta} or \eqref{Flowq} (assuming a viscoplastic approximation) is typically achieved via a predictor-corrector approach for the time-discrete problem. In the predictor step the dissipation function is approximated by a smooth function and the resulting minimization problem solved for $\bu$ and intermediate values of $\gamma^\alpha$ and $\eta^\alpha$. Then, in the corrector step \eqref{eq:phieta} is solved for $\gamma^\alpha$, using $\bu$ to update the stress. 

\section{Stress-gradient plasticity} 
In this section we extend the framework presented earlier, to a model of stress-gradient plasticity. In \cite{chakravarthy-curtin2011} a stress gradient enhancement is proposed and its properties explored. Attention is not however given to the thermodynamic consistency of that model: the authors state: ``It is important to note that such a formulation neglects the formal need for higher-order work terms involving a strain-like variable that would be work-conjugate to the stress gradient and for the accompanying higher-order boundary conditions. Development of
such a full theory is beyond the scope of the present paper." 

The objective of this work is to propose a model of stress-gradient plasticity that is mechanically relevant and thermodynamically consistent. Motivated by the model in \cite{chakravarthy-curtin2011}, we propose an extension of the conventional yield function \eqref{eq:yieldalpha} of the form, for slip system $\alpha$,
\begin{equation}
\phi = |\tau^\alpha| - \big[\tau_0 + \ell f^\alpha (\nabla\vec{\tau})\big]\,,
\label{yield_stressgrad_short}
\end{equation}
where $\ell$ is a length scale, $\nabla\vec{\tau} = \{ \nabla\tau^\beta\ |\ \beta = 1,\ldots, A\}$ and $f^\alpha$ a function to be specified. For the purpose of this study we select $f^\alpha (\nabla\vec{\tau} ) = \sum_\beta |\nabla\tau^\beta\cdot\bs^\alpha |$, so that the yield function is now 
\begin{equation}
\phi = |\tau^\alpha| - \Big[\tau_0 + \ell\sum_\beta |\nabla\tau^\beta\cdot\bs^\alpha |\Big]\,.
\label{yield_stressgrad}
\end{equation}
One may compare the term for current yield, that is, $\tau_0 + \ell\sum_\beta |\nabla\tau^\beta\cdot\bs^\alpha|$, with that in \cite{chakravarthy-curtin2011}, which is of the form
\begin{equation}
\sigma_Y' = \displaystyle \frac{\sigma_Y}{1 - \displaystyle\frac{\ell|\nabla\sigma_e|f(\varepsilon^p_e)}{\sigma_e}}\,.
\label{CCyield}
\end{equation}
Here $\sigma_Y$ and $\sigma_Y'$ are respectively the conventional initial yield stress and the stress-gradient-dependent yield, $\sigma_e$ is a scalar equivalent stress, and $f$ is a hardening term that depends on equivalent plastic strain $\varepsilon^p_e$. Setting aside the hardening term, to first order the expression for $\sigma_Y'$ may be approximated as
\begin{equation}
\sigma_Y' \simeq \sigma_Y + \ell\frac{\sigma_Y}{\sigma_e}  |\nabla\sigma_e|\,.
\label{CCyieldapprox}
\end{equation}
The similarity to the current yield term in \eqref{yield_stressgrad} is evident.

The model presented here is a phenomenological extension of the dislocation-based model developed in \cite{chakravarthy-curtin2011}, in which the length scale is related to obstacle spacing. The current theory thus captures in an averaged way the key elements of that model, and the length scale does not have a direct physical interpretation. This is similar to the situation for phenomenological strain-gradient theories. 

Now define
\begin{equation}
g^{\alpha\beta} := \nabla\tau^\beta\cdot\bs^\alpha
\label{galphabeta}
\end{equation}
and 
\begin{equation}
g^{\alpha} = - \sum_\beta |g^{\alpha\beta}|\,.
\label{galpha}
\end{equation}
Then the yield function is
\begin{equation}
\phi (\tau^\alpha,g^\alpha) = |\tau^\alpha| - [\tau_0 - \ell g^\alpha]\,.
\label{yield_stressgrad_new}
\end{equation}
Proceeding as in Section \ref{rate-ind}, the normality law again gives
\begin{subequations}
\begin{align}
\dot{\gamma}^\alpha & = \lambda^\alpha \frac{\partial \phi}{\partial \tau^\alpha}\,,  
\label{eq:normality_a}\\
\dot{\eta}^\alpha & = \lambda^\alpha \frac{\partial \phi}{\partial g^\alpha} = \lambda^\alpha\ell\,.
\label{eq:normality_b}
\end{align}
\label{eq:normality_stressgrad}
\end{subequations}
Next we define the free energy as before, that is,
\begin{equation}
\psi = \psi^e (\be) + \psi^h (\vec{\eta})\,,
\label{eq:psi_stress}
\end{equation}
with the thermodynamic conjugate given by
\begin{equation}
g^\alpha = -\frac{\partial \psi^h}{\partial \eta^\alpha}\,.
\label{galpha_stress}
\end{equation}
Equations \eqref{galphabeta}, \eqref{galpha} and \eqref{galpha_stress} give a relationship between $\eta^\alpha$ and the stress, that is,
\begin{equation}
g^\alpha(\vec{\eta}) = - \sum_\beta|\nabla\tau^\beta\cdot\bs^\alpha|\,.
\label{g_ito_eta}
\end{equation}
Equations \eqref{g_ito_eta} constitute a set of equations that may be inverted to give $\eta^\alpha$ in terms of the resolved stress gradients. Unlike the classical case, however, we do not specify $\psi^h (\vec{\eta})$ explicitly, nor will we need to do so as $\eta^\alpha$ do not enter the solution procedure. 

{\bf The flow relation in terms of the dissipation.}\ \ All of the details in Section \ref{rate-ind} concerning the dissipation function carry over unchanged to the model of stress-gradient plasticity considered here. In the classical case, for the flow relation in terms of $D_{\rm eff}$, that is, \eqref{eq:flowDeffSC}, \eqref{eq:flowDeffSC2}, or \eqref{Flowq}, one substitutes directly for $g^\alpha$ in terms of the equivalent plastic strain $\eta^\alpha$. 
For the stress-gradient case we have the same flow relation, with $g^\alpha$ defined by \eqref{galpha}.

{\bf A weak formulation for $g^{\alpha\beta}$.}\ \ We can obtain $g^\alpha$ without the need to compute derivatives of the stress as in \eqref{galphabeta}, as follows. We start by constructing a weak form for $g^{\alpha\beta}$ using \eqref{galphabeta}. For this purpose 
introduce a test function $w \in H^1(\Omega)$, multiply both sides of \eqref{galphabeta} by $w$ and integrate, then integrate by parts to obtain 
\begin{align}
\int_\Omega g^{\alpha\beta}w\ dx & =  \int_{\Omega} [\nabla\tau^\beta\cdot\bs^\alpha]w\ dx \nonumber \\
& =  \int_{\partial \Omega} [\tau^\beta\bs^\alpha\cdot\bn] w\ ds - \int_{\Omega} \tau^\beta\bs^\alpha\cdot\nabla w\ dx\,.  
\label{weakG}
\end{align}
This is an equation for $g^{\alpha\beta}$, for given $\tau^\alpha$, and $g^\alpha$ is then obtained from \eqref{galpha}. The weak formulation allows for the right-hand side to be computed without having to approximate the gradients of the resolved shear stress.

\subsection{A solution algorithm}
The complete formulation of the problem for stress-gradient plasticity comprises the equilbrium equation \eqref{weak_equil}, the flow relation \eqref{eq:phieta} or \eqref{Flowq}, and \eqref{weakG}: three equations for $\bu, \vec{\gamma}, \vec{g}$, with the resolved shear stress $\tau^\alpha$ found from \eqref{eq:resshear1} and the elastic relation \eqref{eq:elastreln} in terms of $\bu$ and $\vec{\gamma}$.

{\bf The discrete equation for $g^{\alpha\beta}$.}\ \ Consider a finite element formulation comprising a mesh of linear ($P_1$) triangles, and set 
\begin{equation}
w = {\sf Nw},\qquad g^{\alpha\beta} = {\sf Ng^{\alpha\beta}},\qquad \nabla w = {\sf Bw}\,,
\label{GFE}
\end{equation}
where ${\sf N}$ is the vector of shape functions, ${\sf B}$ the vector of shape function derivatives, and ${\sf w}$ and ${\sf g}^{\alpha\beta}$ are vectors of nodal degrees of freedom. Substitute in \eqref{weakG} to get
\begin{equation}
{\sf Mg}^{\alpha\beta} = \int_{\partial\Omega} [ \tau^\beta\bs^\alpha\cdot\bn]\,{\sf N^T}\ ds - \int_\Omega {\sf B^T}\bs^\alpha \tau^\beta\ dx\,,
\label{discreteG}
\end{equation}
where ${\sf M}$ is the standard mass matrix. This may be solved for $g^{\alpha\beta}$, and hence $g^\alpha$, for a given stress.

{\em Algorithm.}\ \ 
\begin{enumerate}
\item[1.] For the incremental problem, assume that everything is known a time step $t_{n}$.
\item[2.]
Solve the equilibrium equation \eqref{weak_equil} for $\boldmath u_{n+1}$ and evaluate the predictor stress using $\tau^{\alpha\,{\rm pred}}_{n+1} = \tau^\alpha_{n+1}(\boldmath u_{n+1},g^\alpha_n)$.
\item[3.]
Find $g^{\alpha\beta}_{n+1}$ from \eqref{discreteG}, and $g^\alpha_{n+1}$ from \eqref{galpha}.
\item[4.]
Update the slip increments by using \eqref{eq:phieta}.
\end{enumerate}

The problem \eqref{weakG} and associated algorithm are well posed; one solves \eqref{discreteG} for a given stress distribution, and finite element approximations converge even for discontinuous stress distributions.  

\section{An example: torsion of a thin wire}
We consider the problem studied in \cite{Steinmann_etal2019} of a single crystal cylindrical specimen of radius $R = 10\,\mu m$, with face centred cubic unit cells is subjected to simple torsion in the form of a constant twist $\kappa$ per unit length.
There is no dependence of the response on the axial coordinate $z$, which coincides with the cylinder axis, and it suffices to consider a section of length $L$. The boundary conditions relative to a cylindrical coordinate system are
\begin{subequations}
\begin{align}
& u_\theta(r,\theta,0) = u_z(r,\theta,0) = u_z(r,\theta,L) = 0\,,
\quad u_\theta(r,\theta,L) = \kappa L\,,
\label{disp_bc}\\
& t_r (r,\theta,0) = t_r (r,\theta,L) = 0\,, \quad \bt (R,\theta,z) = \bzero\,.
\label{tracn_bc}
\end{align}
\label{torsion_bcs}
\end{subequations}
These conditions are applied to the equilibrium equation \eqref{weak_equil}. With regard to the evaluation of $g^{\alpha\beta}$ from \eqref{discreteG}, the boundary term when expanded reads 
\[
\int_0^z\int_0^{2\pi} F(r,\theta)\, \bn\ rd\theta dz + 
\int_0^R\int_0^{2\pi} F(r,\theta) (-\be_z)\ rdrd\theta + 
\int_0^R\int_0^{2\pi} F(r,\theta)\, \be_z\ rdrd\theta  \,.
\]  
Here we have denoted the integrand by $F$ and indicated explicitly its independence of $z$. The three terms correspond to the integrals respectively over the curved boundary, the end $z = 0$ and the end $z = L$. The normals to these two latter boundaries are respectively $\mp\be_z$. It follows that the second and third terms cancel, leaving the boundary integral 
\[
\int_0^z\int_0^{2\pi} F(r,\theta)\, \bn\ rd\theta dz\quad \mbox{or}\quad  L\int_0^{2\pi} F(r,\theta)\, \bn\ rd\theta \,.
\]
The components of the elasticity tensor are given by \eqref{elastcubic}, and the material properties \eqref{eq:coeffCubSym}
are taken from \cite{alankar2009}. Hardening is neglected and the critical Schmid stress $\tau_0$ is set at approximately $10^{-3} \times$ Young's modulus. The relevant parameters are given in Table \ref{tab:parameters}, while the length scale $\ell$ and the self/latent ratio $q$ in \eqref{gchi} are varied and given in the result plots. 

\begin{figure}[ht!]
\centering
\inputTikz{tikzPics/AK_Unitcell.tex}{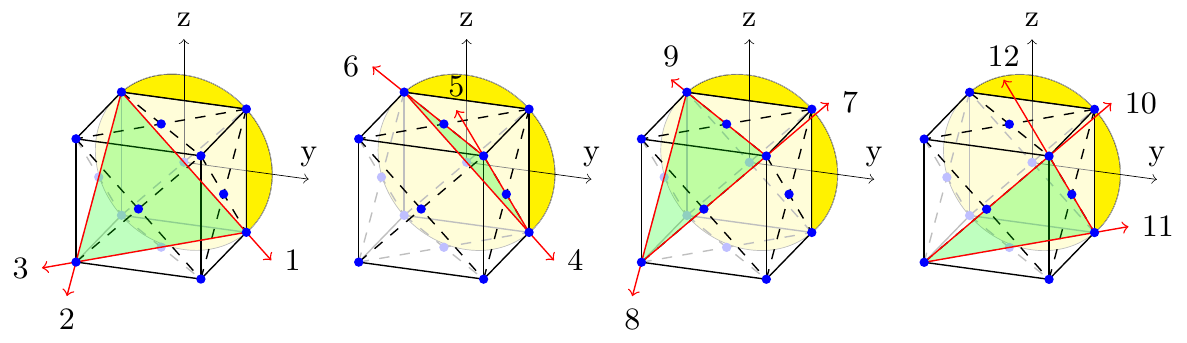}
\caption{Wire cross section shown in yellow with the face centred unit cell and one of the four slip planes shaded in green in each subfigure. The three edges of the slip planes represent the crystal slip directions and thus the slip directions $\bs^\alpha$.}
\label{fig:unitcellInWire}
\end{figure}

\begin{figure}[!ht]
  \includegraphics[height = 0.3\textwidth]{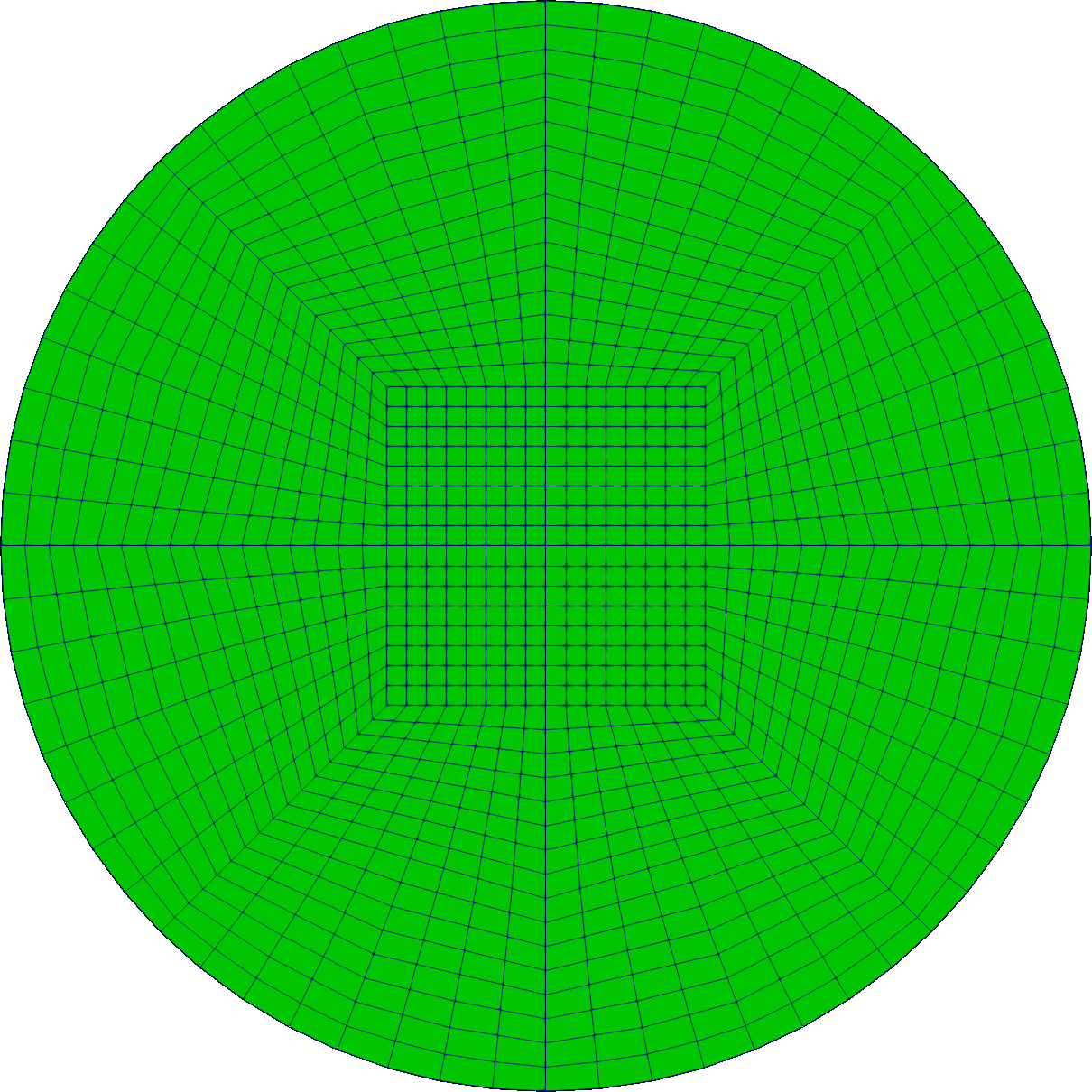}
  \hspace{1cm}
  \includegraphics[height = 0.3\textwidth]{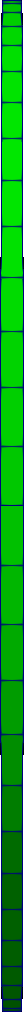}
   \centering
 \caption{Section of the wire showing a mesh comprising 1280 tri-linear hexahedral elements}
 \label{fig:mesh}
\end{figure}

The total load of $\kappa = 5\,mm^{-1}$ is applied in 30 equidistant time steps with a rate of $40\,mm^{-1} s^{-1}$. The orientation of the unit cells relative to the wire is depicted in Figure \ref{fig:unitcellInWire}. For an easier interpretation of the results
the unit cells are aligned with the wire axis. We do not vary the orientation as  this is not a focus of the study.
A mesh containing 1280 tri-linear elements for the displacements is depicted in Figure \ref{fig:mesh}.

\begin{table}[!ht]
 \begin{center}
  \begin{tabular}{|ll|ll|ll|ll|ll|}
  \hline
   $E$				&	{63.6} {\rm GPa}&
   $\mu$			&	{28.5} {\rm GPa}&
   $\nu$			&	{0.362} {\rm GPa}&
   $\tau_0$			&	{60} {\rm MPa}&
   $p'$ (see \eqref{Flowq})				&	21\\
  \hline
  \end{tabular}
\caption{Material parameters for the model}
      \label{tab:parameters}
 \end{center}
\end{table}

\begin{figure}[ht!]
\centering
\inputTikz{TikzPics/AK_diff_l_r.tex}{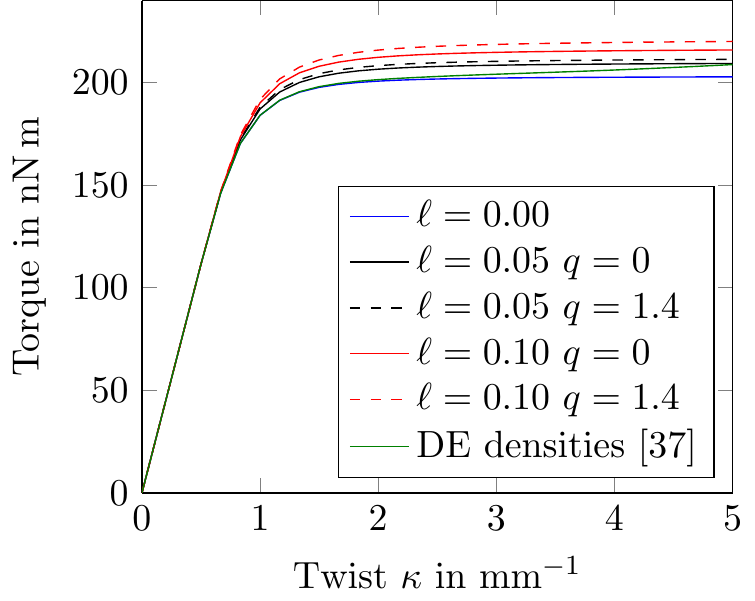}
\caption{Torque vs. twist for different values of $\ell$ in \si{\micro\meter} and $q$ compared with results from the disequilibrium density (DE) model introduced in \cite{Steinmann_etal2019}. }
\label{fig:diff_l_r}
\end{figure}

\begin{figure}[ht!]
\centering
\inputTikz{TikzPics/AK_diff_l_r_zoom.tex}{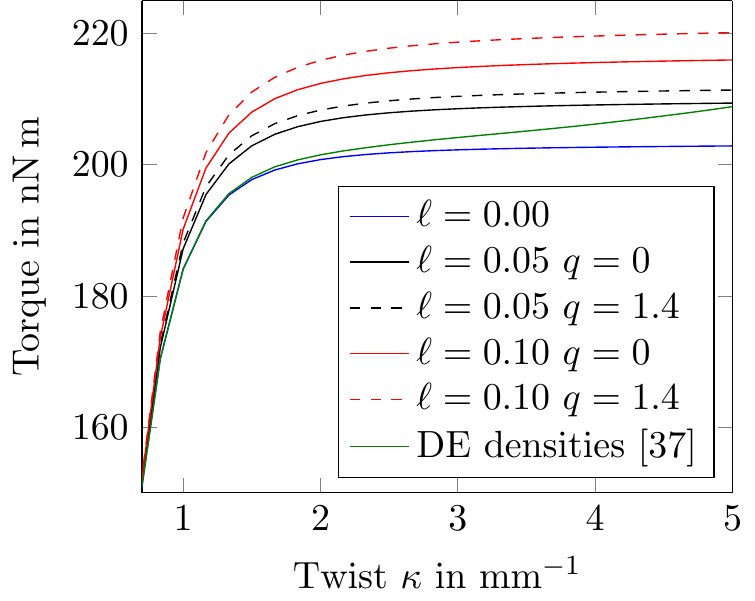}
\caption{Torque vs. twist for different values of $\ell$ in \si{\micro\meter} and $q$ as in Figure \ref{fig:diff_l_r}, magnified}
\label{fig:diff_l_r_zoom}
\end{figure}

The torque-twist curves for different values of the stress-gradient length scale $\ell$ and the interaction coefficient $q$ are shown in Figure \ref{fig:diff_l_r}, and in magnified form in Figure \ref{fig:diff_l_r_zoom}. The results obtained using the disequilibrium density formulation introduced in \cite{Steinmann_etal2019} are also shown.  

Table \ref{tab:torque} gives the limiting torque, defined as $T_l = T(\kappa = 5 mm^{-1})$ and 
$q=1.0$. The strengthening, that is, the increase in limiting torque with length scale, is clearly seen. Since no other hardening is considered the limiting torque $T_l$ correlates with the torque at the initialization of plastic deformation $T_y$, and from Table \ref{tab:torque} the relationship between limiting torque and length scale is approximately linear. The same holds for the relationship with the self/latent ratio $q$. Hardening behaviour, that is, an increase in torque with increase in twist, is mild. This suggests that the stress gradients are close to being constant, resulting in minimal increase in the current yield torque. In contrast, the disequilibrium densities lead to further hardening, since gradients of plastic quantities are considered in the critical Schmid stress.

Another ansatz for the inclusion of gradients of plastic quantities within a crystal plasticity model would be the incorporation of dislocation densities as in \cite{Bardella-Panteghini2015, Steinmann_etal2019}, for example. From a macroscopic point of view this leads to size dependent hardening, similar to that for the disequilibrium densities model in Figure \ref{fig:diff_l_r} and \ref{fig:diff_l_r_zoom}. A detailed comparison is given in \cite{Steinmann_etal2019}. By way of comparison the results for the polycrystalline case investigated in \cite{Idiart-Fleck2010} using a strain gradient theory show similar size-dependent strengthening to that presented here, albeit accompanied by slight hardening, arising from the inclusion in that model of a hardening term that depends on a measure of plastic strain and its gradient.

\begin{figure}[!ht]
\captionsetup[subfigure]{labelformat=empty}
\centering
\ownCBar{0.4\textwidth}{Von Mises stress}{22.83}{207.77}{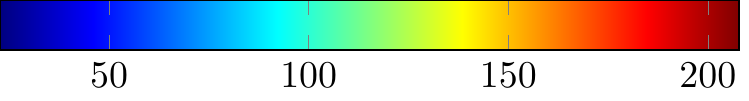}{50,100,150,200}
\\
\vspace{1mm}
\centering
  \begin{subfigure}[t]{0.28\textwidth}
    \includegraphics[width = \textwidth]{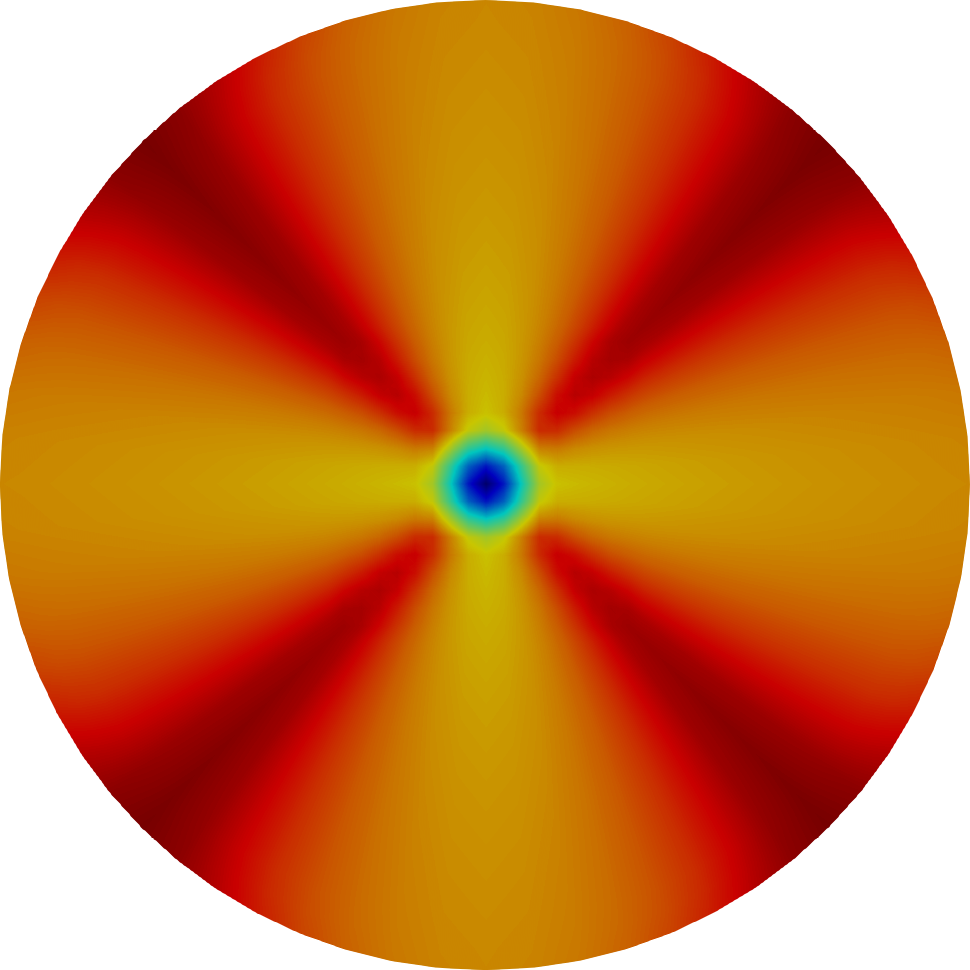}
    \caption{\small $\ell = 0.0$}
    \label{fig:Mises:l0_r0}
  \end{subfigure}
    \centering \hspace{0.01\textwidth}
  \begin{subfigure}[t]{0.28\textwidth}
    \includegraphics[width = \textwidth]{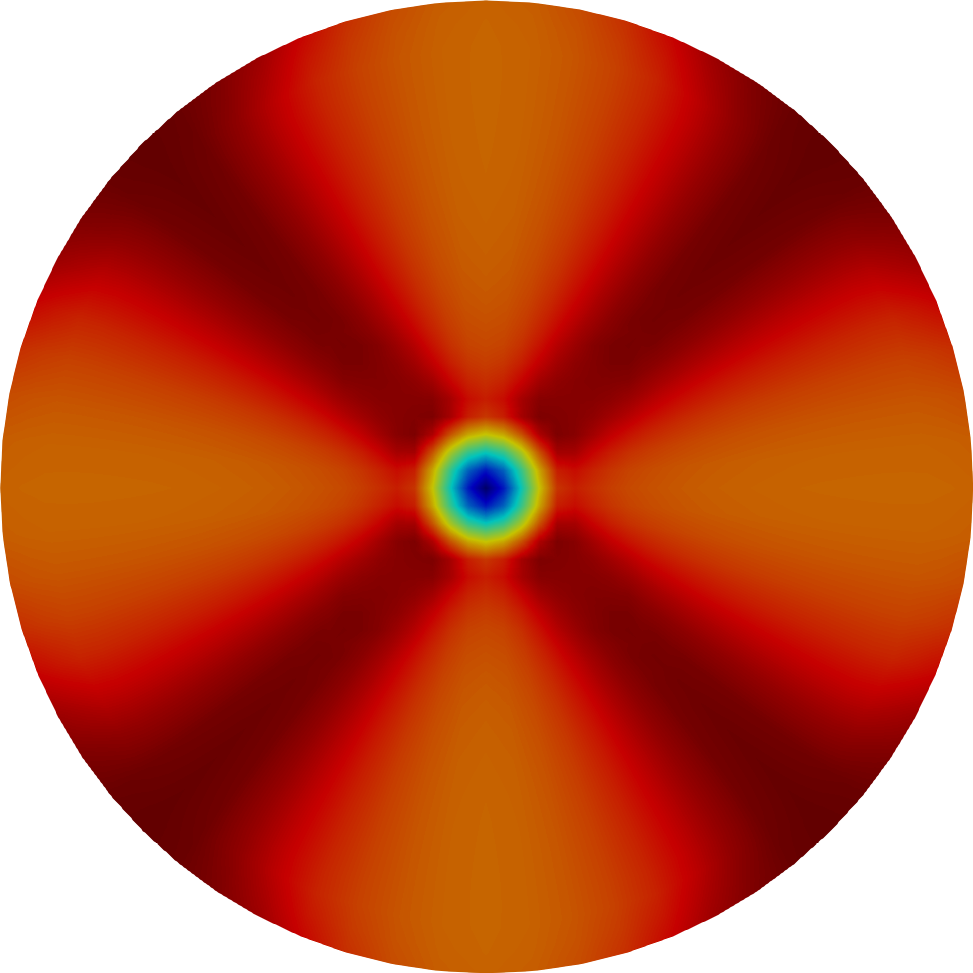}
    \caption{\small $\ell = 0.05; \, q = 0.0$}
    \label{fig:Mises:l005_r0}
  \end{subfigure}
    \centering \hspace{0.01\textwidth}
  \begin{subfigure}[t]{0.28\textwidth}
    \includegraphics[width = \textwidth]{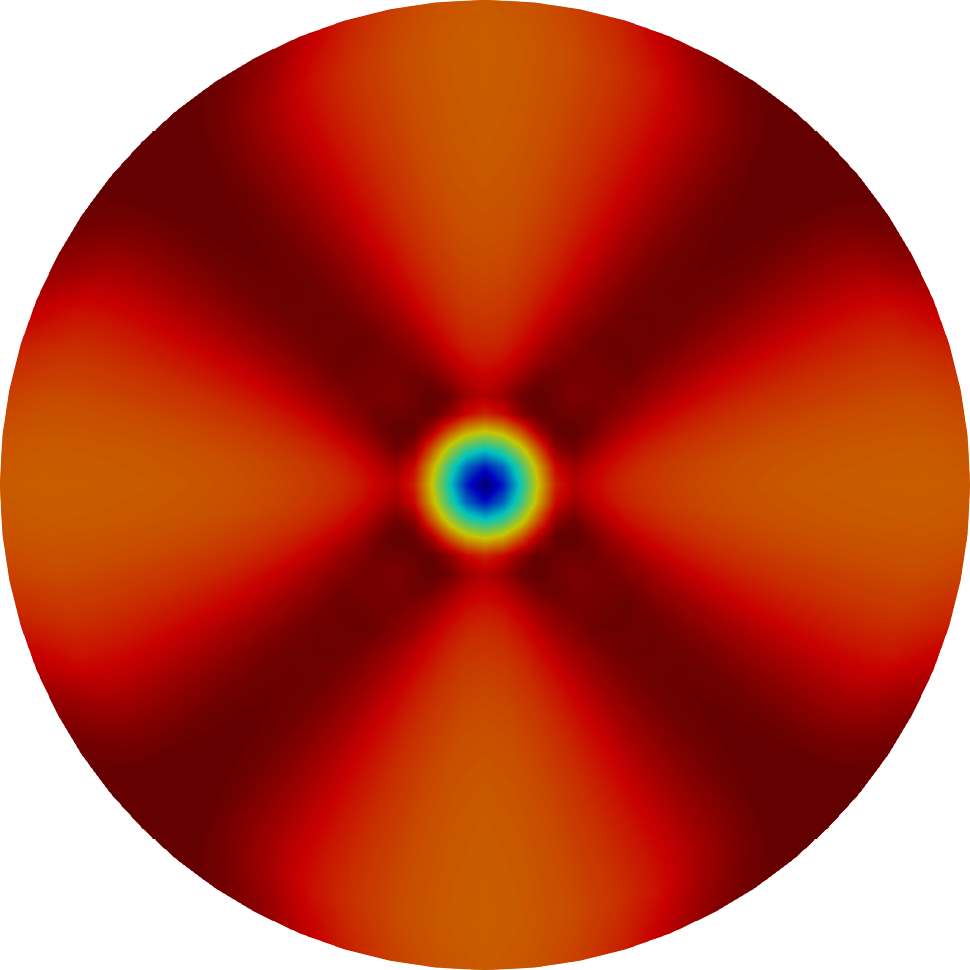}
    \caption{\small $\ell = 0.05; \, q = 1.4$}
    \label{fig:Mises:l005_r14}
  \end{subfigure}
 \caption{Distribution of von Mises stress (MPa) across wire section at twist of {5}{\rm mm}$^{-1}$}
 \label{fig:Mises}
\end{figure}

\begin{figure}[!ht]
\captionsetup[subfigure]{labelformat=empty}
\centering
\ownCBar{0.4\textwidth}{Equivalent plastic strain}{0}{0.0209}{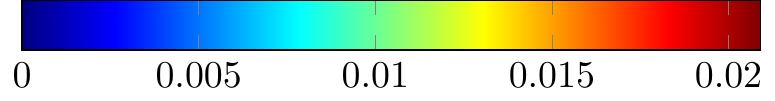}{0, 0.005, 0.01, 0.015, 0.02}
\\
\vspace{1mm}
\centering
  \begin{subfigure}[t]{0.28\textwidth}
    \includegraphics[width = \textwidth]{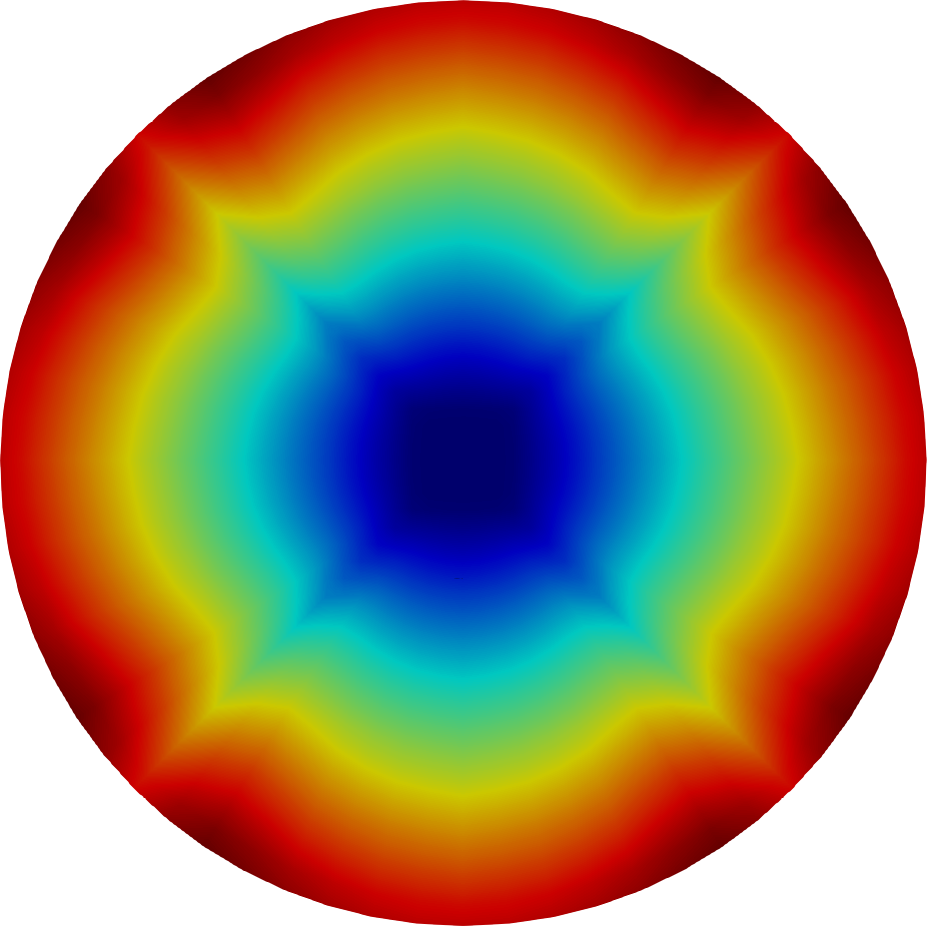}
    \caption{\small $\ell = 0.0$}
    \label{fig:equivPlStrain:l0_r0}
  \end{subfigure}
    \centering \hspace{0.01\textwidth}
  \begin{subfigure}[t]{0.28\textwidth}
    \includegraphics[width = \textwidth]{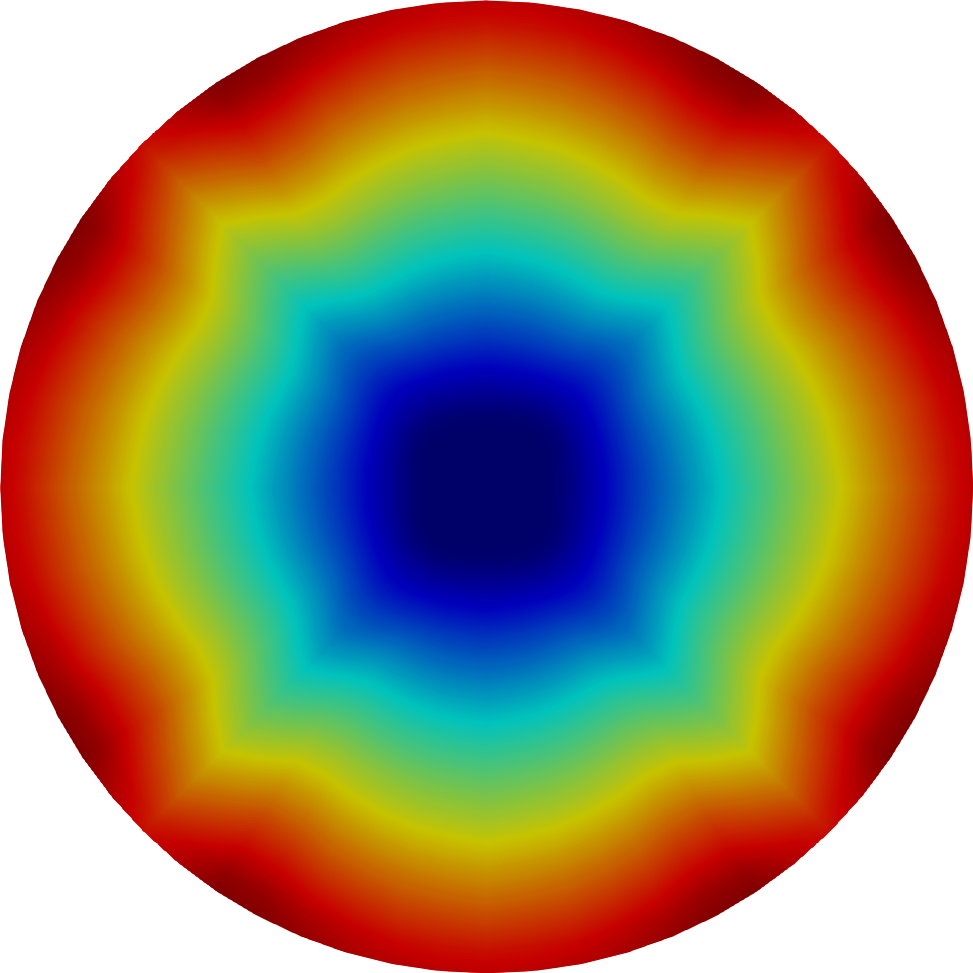}
    \caption{\small $\ell = 0.05; \, q = 0.0$}
    \label{fig:equivPlStrain:l005_r0}
  \end{subfigure}
    \centering \hspace{0.01\textwidth}
  \begin{subfigure}[t]{0.28\textwidth}
    \includegraphics[width = \textwidth]{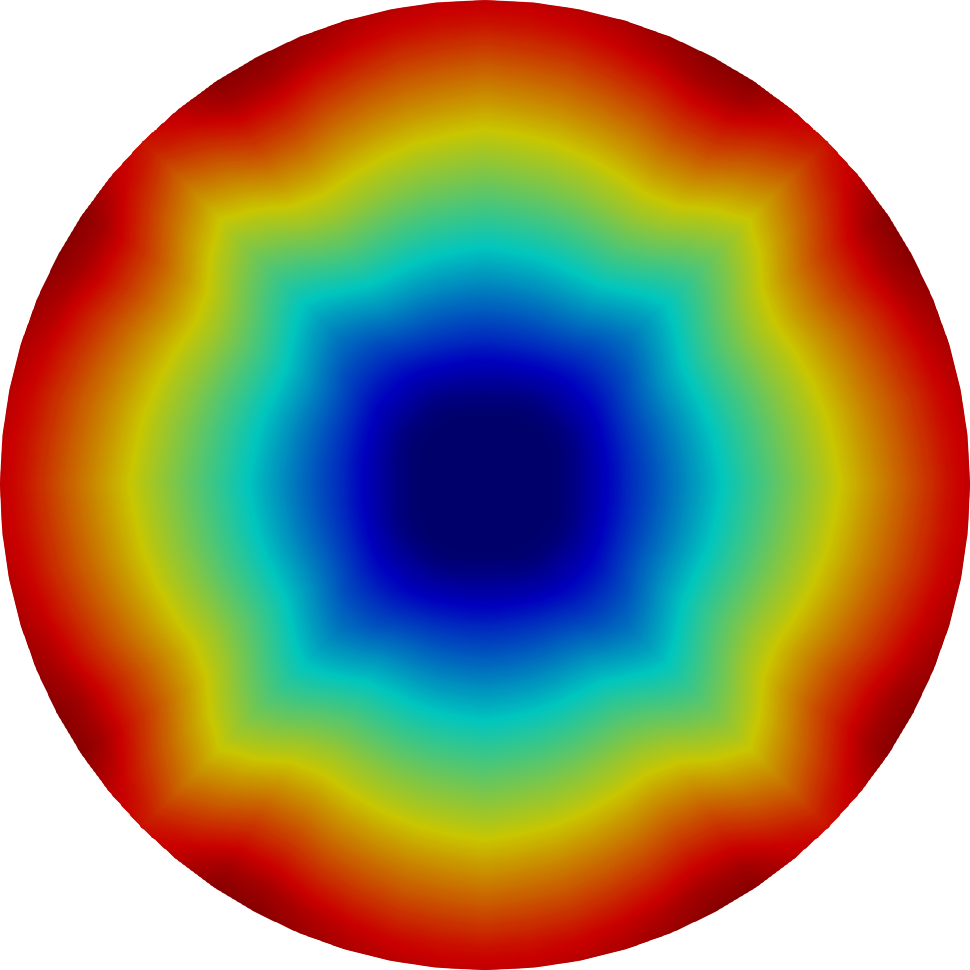}
    \caption{\small $\ell = 0.05; \, q = 1.4$}
    \label{fig:equivPlStrain:l005_r14}
  \end{subfigure}
 \caption{Distribution of equivalent plastic strain at twist of {5}{\rm mm}$^{-1}$}
 \label{fig:equivPlStrain}
\end{figure}

Figures \ref{fig:Mises} and \ref{fig:equivPlStrain} show the spatial distribution of von Mises stress and equivalent plastic strain in the wire cross section for different values of $q$ and $\ell$. It can be seen that for the conventional theory ($\ell = 0$) the stresses, and as a result the plastic strains, are concentrated along diagonals of the wire. These diagonals coincide with the edges of the crystal unit cell. The stress and plastic strain distributions become more diffuse as the values of  $\ell$ and $q$ increase in combination. 

\begin{figure}[!ht]
\captionsetup[subfigure]{labelformat=empty}
\centering
\ownCBar{0.4\textwidth}{ga}{0.08498523384332657}{20.7932071685791}{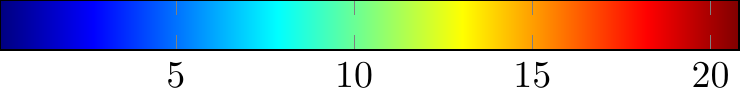}{5,10,15,20}
\\
\vspace{2mm}
\centering
  \begin{subfigure}[t]{0.22\textwidth}
    \includegraphics[width = \textwidth]{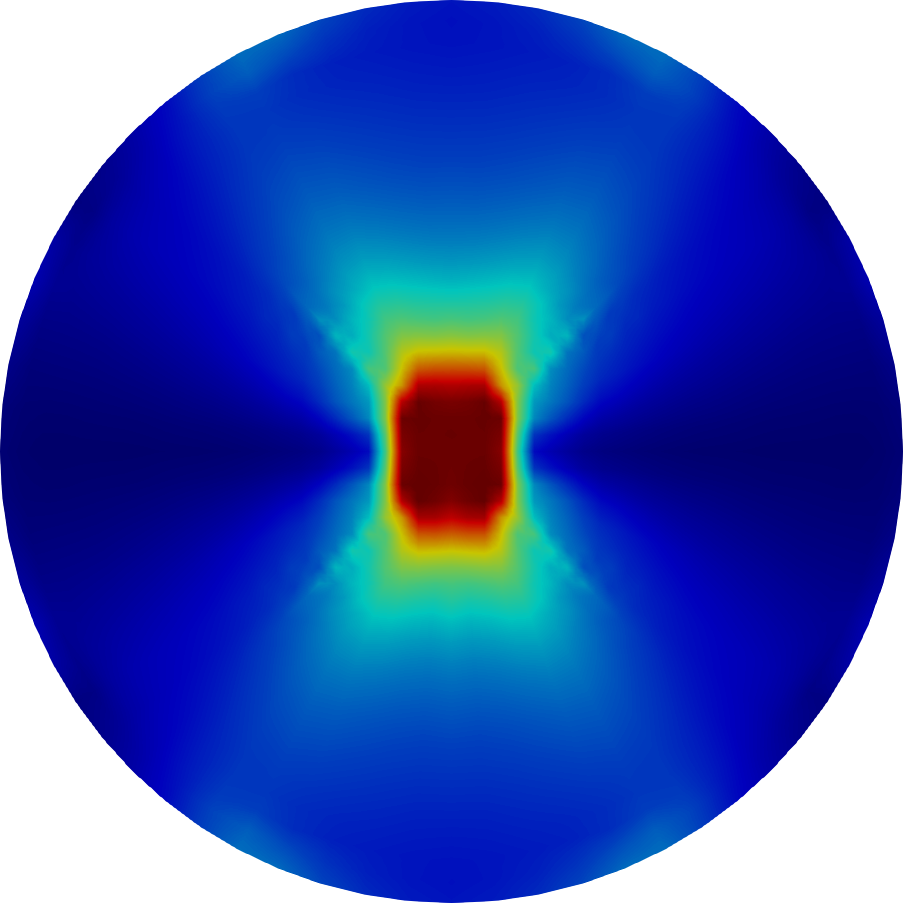}
    \caption{\small $\alpha = 1$}
  \end{subfigure}
  \hspace{0.01\textwidth}
  \begin{subfigure}[t]{0.22\textwidth}
    \includegraphics[width = \textwidth]{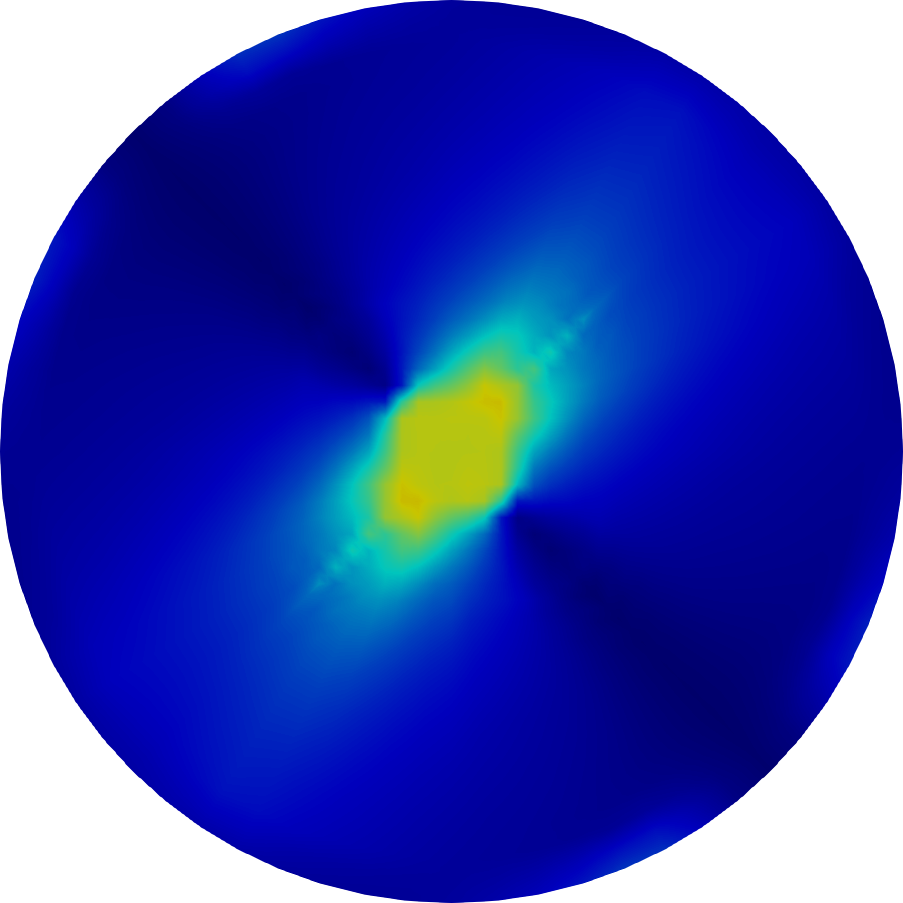}
    \caption{\small $\alpha = 2$}
  \end{subfigure}
  \hspace{0.01\textwidth}
  \begin{subfigure}[t]{0.22\textwidth}
    \includegraphics[width = \textwidth]{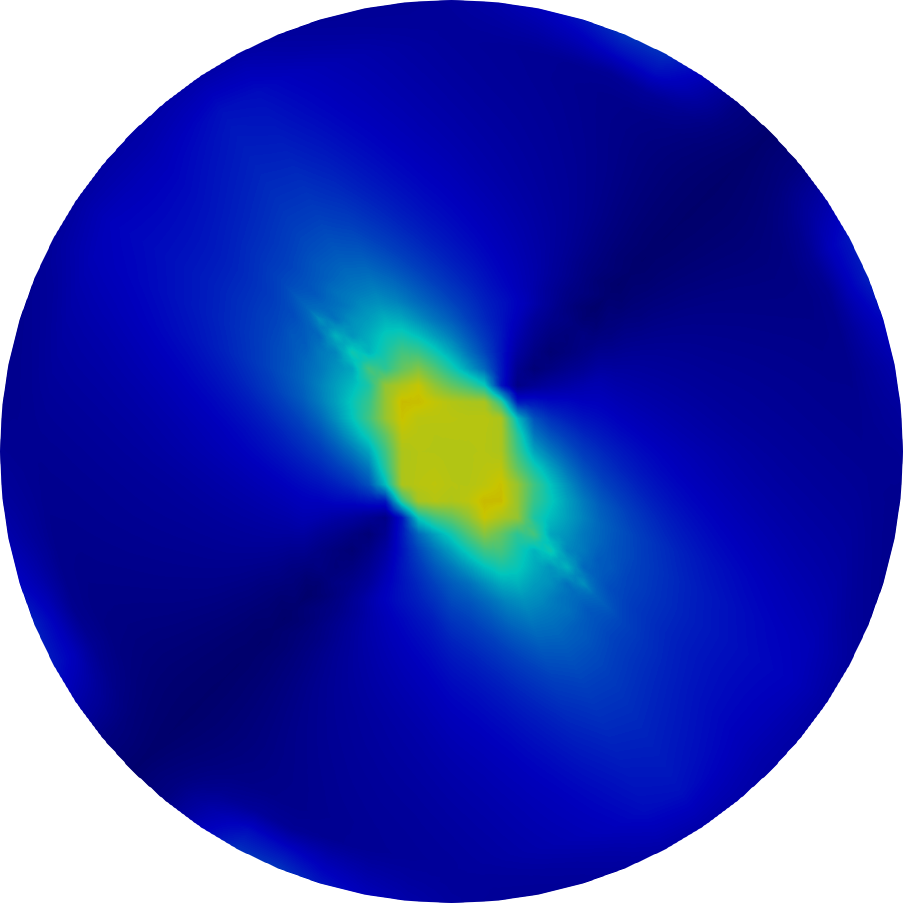}
    \caption{\small $\alpha = 3$}
  \end{subfigure}
  \hspace{0.01\textwidth}
  \begin{subfigure}[t]{0.22\textwidth}
    \includegraphics[width = \textwidth]{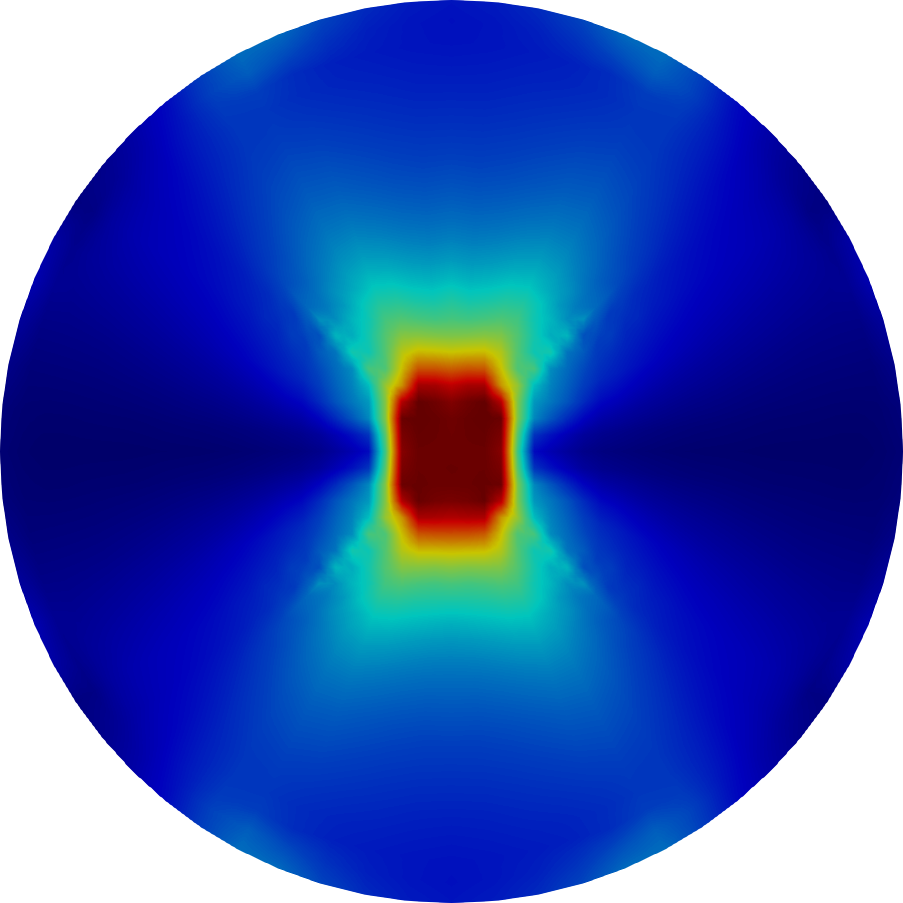}
    \caption{\small $\alpha = 4$}
  \end{subfigure}
  \\
  \vspace{5mm}
  \begin{subfigure}[t]{0.22\textwidth}
    \includegraphics[width = \textwidth]{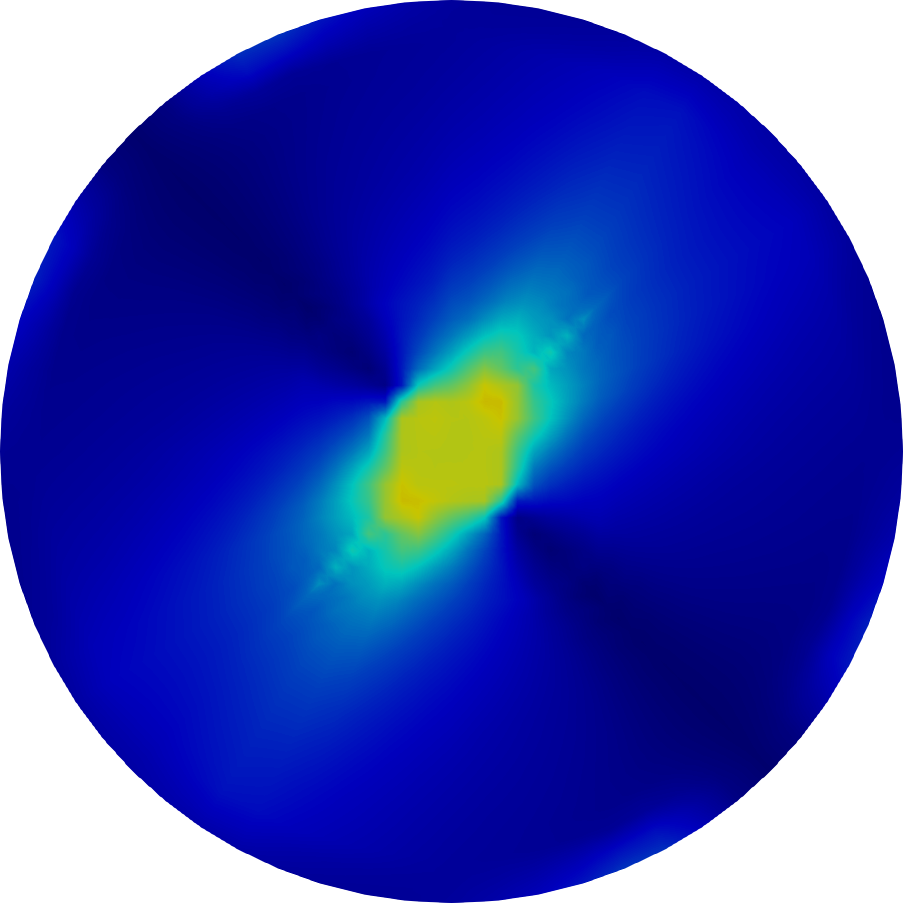}
    \caption{\small $\alpha = 5$}
  \end{subfigure}
  \hspace{0.01\textwidth}
  \begin{subfigure}[t]{0.22\textwidth}
    \includegraphics[width = \textwidth]{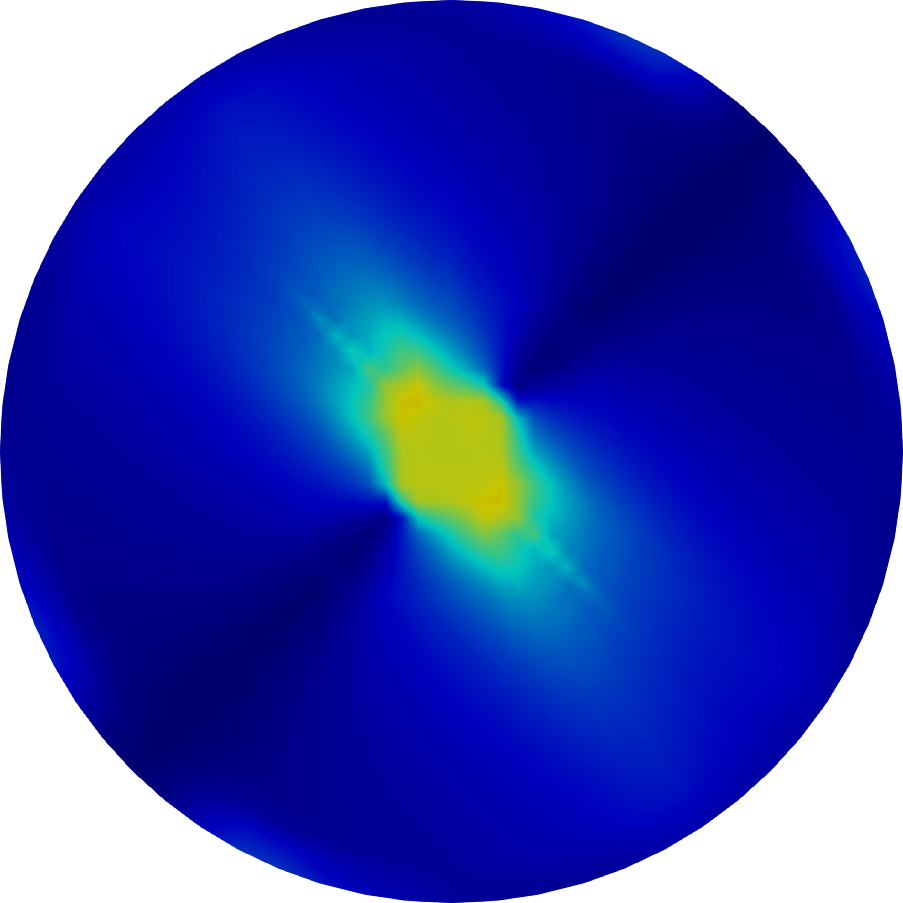}
    \caption{\small $\alpha = 6$}
  \end{subfigure}
  \hspace{0.01\textwidth}
  \begin{subfigure}[t]{0.22\textwidth}
    \includegraphics[width = \textwidth]{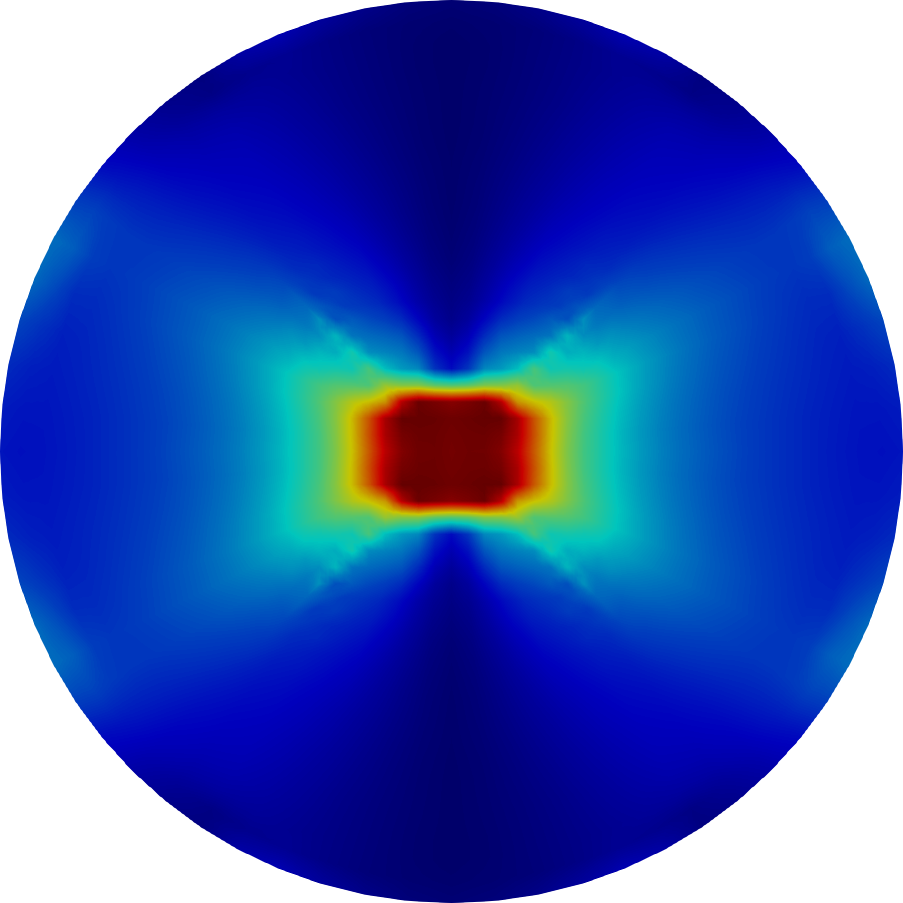}
    \caption{\small $\alpha = 7$}
  \end{subfigure}
  \hspace{0.01\textwidth}
  \begin{subfigure}[t]{0.22\textwidth}
    \includegraphics[width = \textwidth]{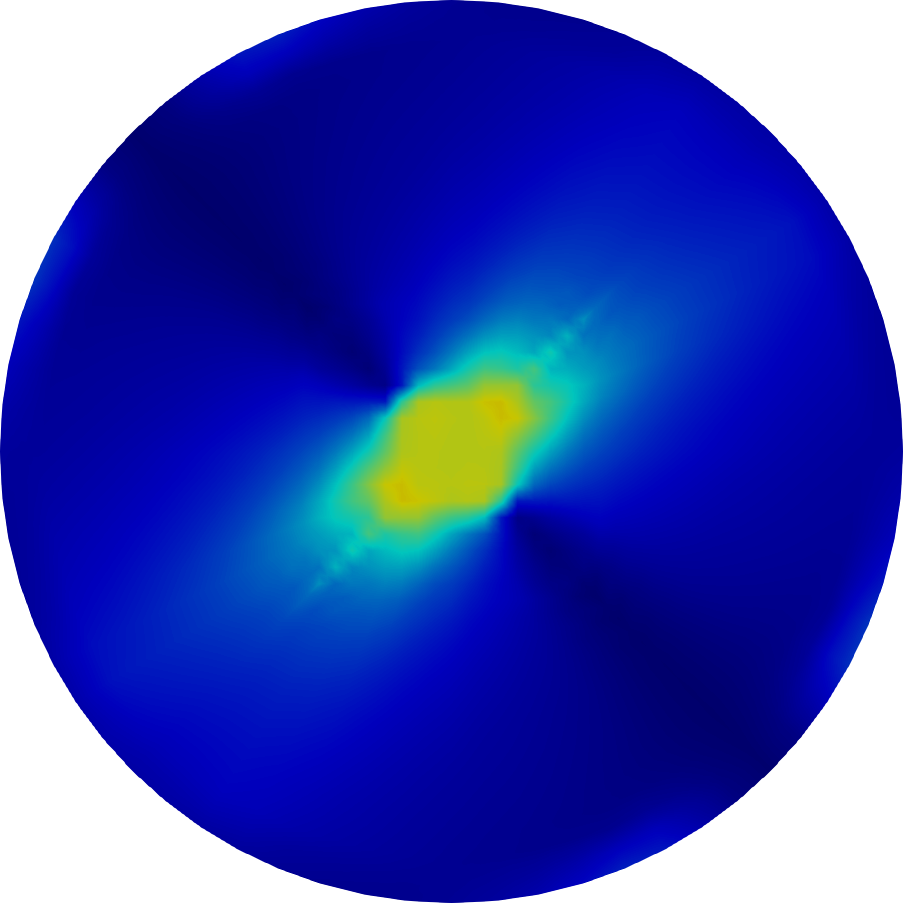}
    \caption{\small $\alpha = 8$}
  \end{subfigure}
  \\
  \vspace{5mm}
  \begin{subfigure}[t]{0.22\textwidth}
    \includegraphics[width = \textwidth]{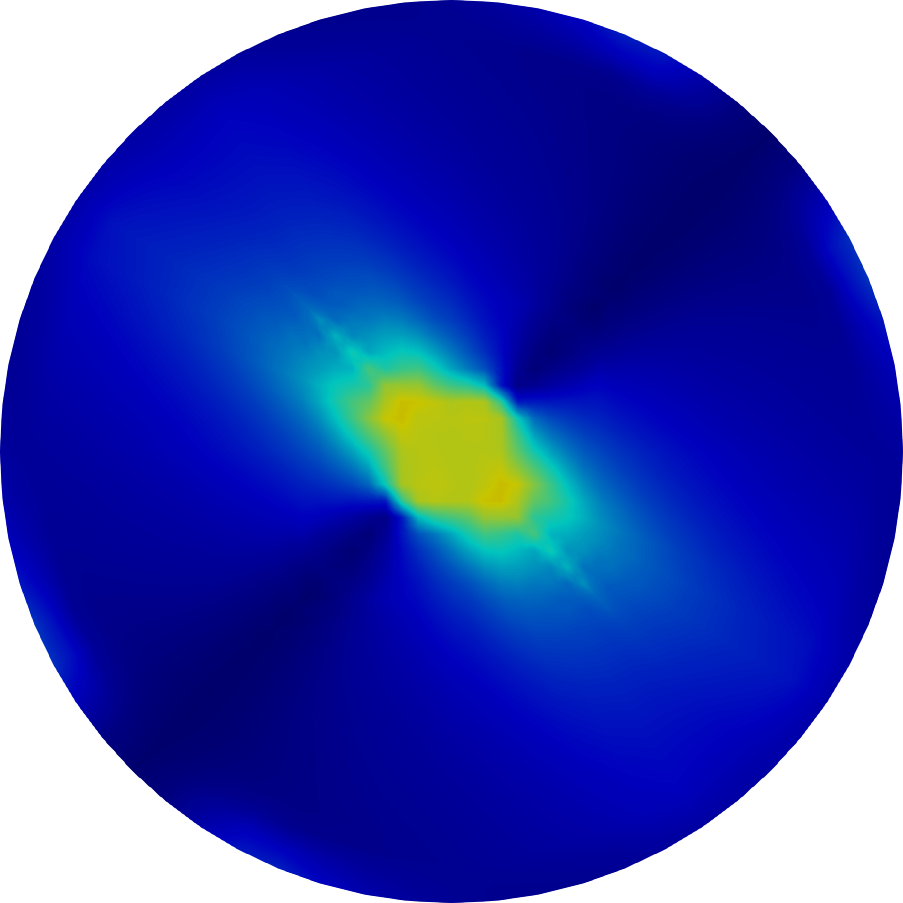}
    \caption{\small $\alpha = 9$}
  \end{subfigure}
  \hspace{0.01\textwidth}
  \begin{subfigure}[t]{0.22\textwidth}
    \includegraphics[width = \textwidth]{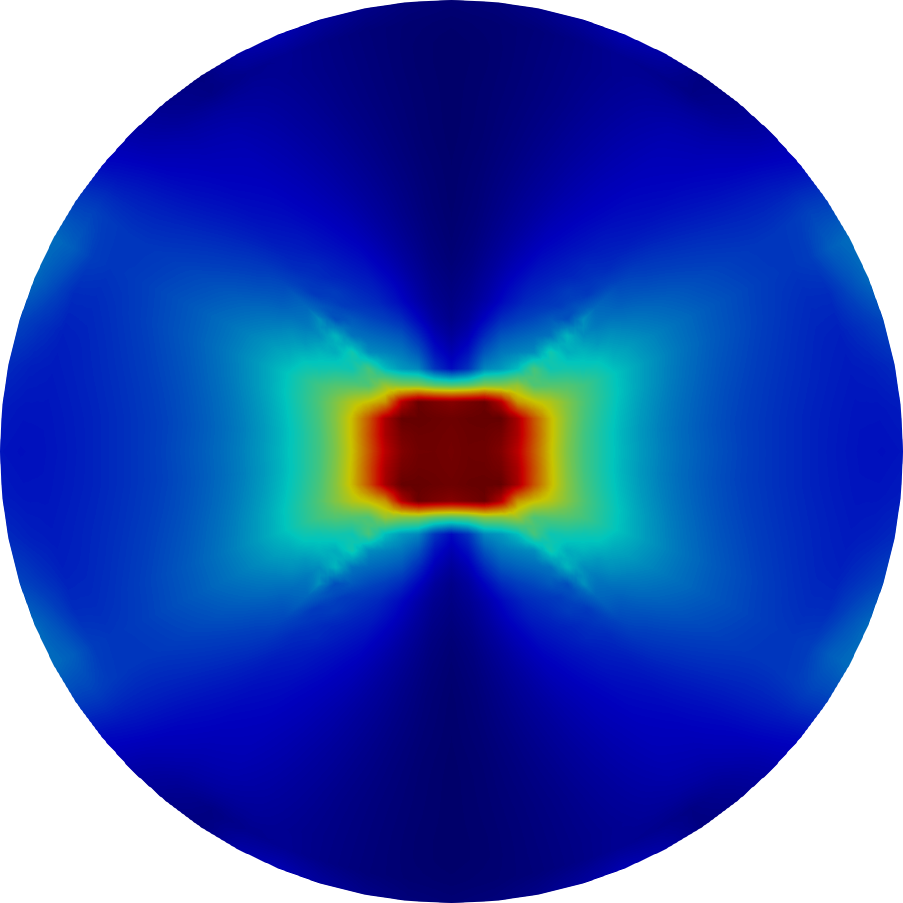}
    \caption{\small $\alpha = 10$}
  \end{subfigure}
  \hspace{0.01\textwidth}
  \begin{subfigure}[t]{0.22\textwidth}
    \includegraphics[width = \textwidth]{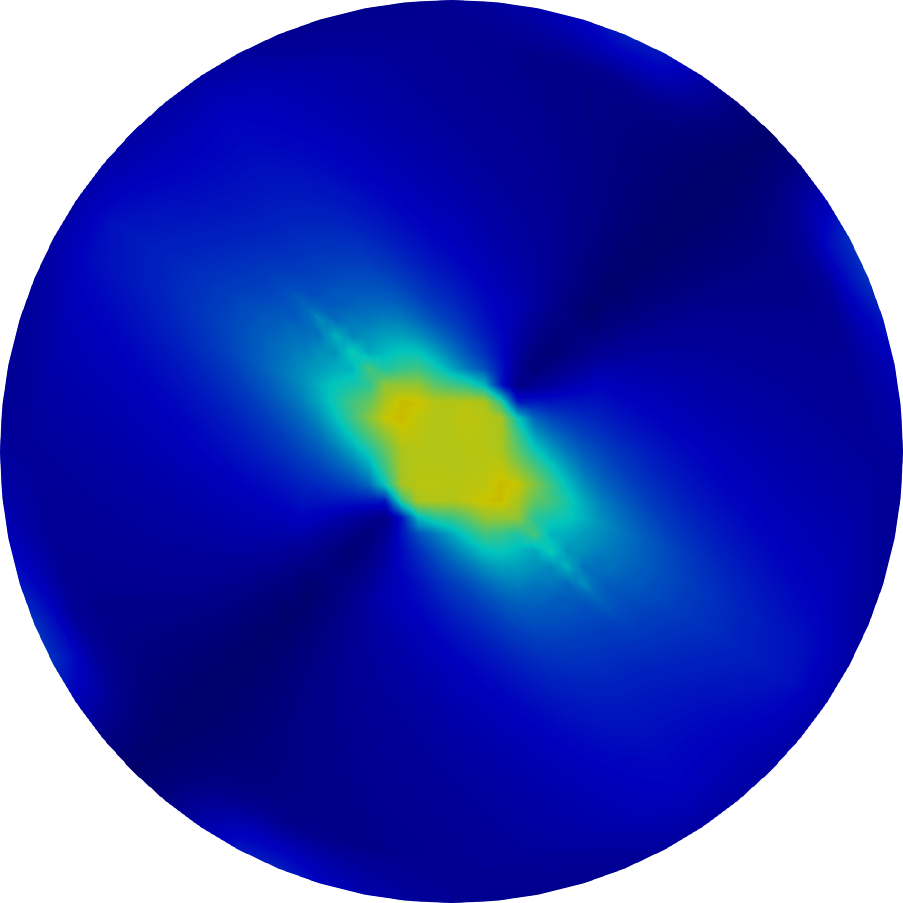}
    \caption{\small $\alpha = 11$}
  \end{subfigure}
  \hspace{0.01\textwidth}
  \begin{subfigure}[t]{0.22\textwidth}
    \includegraphics[width = \textwidth]{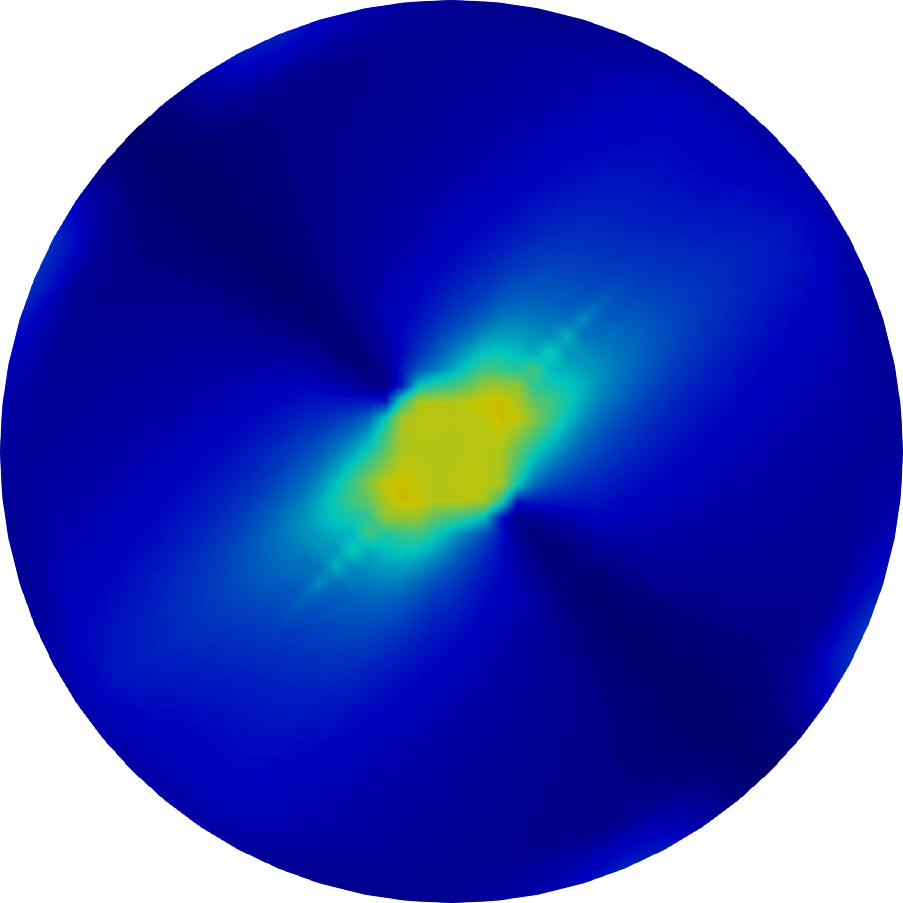}
    \caption{\small $\alpha = 12$}
  \end{subfigure}
  \\
  \vspace{5mm}
 \caption{$g^{\alpha}$ in \si{\mega\pascal} at twist of {5}{\rm mm}$^{-1}$ with $q=0$ and $\ell = 0.05$.}
 \label{fig:ga_l005r0}
\end{figure}

\begin{figure}[!ht]
\captionsetup[subfigure]{labelformat=empty}
\centering
\ownCBar{0.4\textwidth}{ga}{0.1594678908586502}{26.81878662109375}{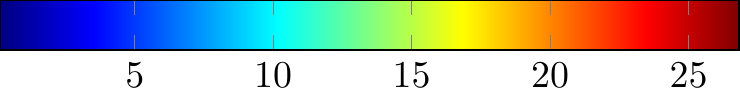}{5,10,15,20,25}
\\
\vspace{2mm}
\centering
  \begin{subfigure}[t]{0.22\textwidth}
    \includegraphics[width = \textwidth]{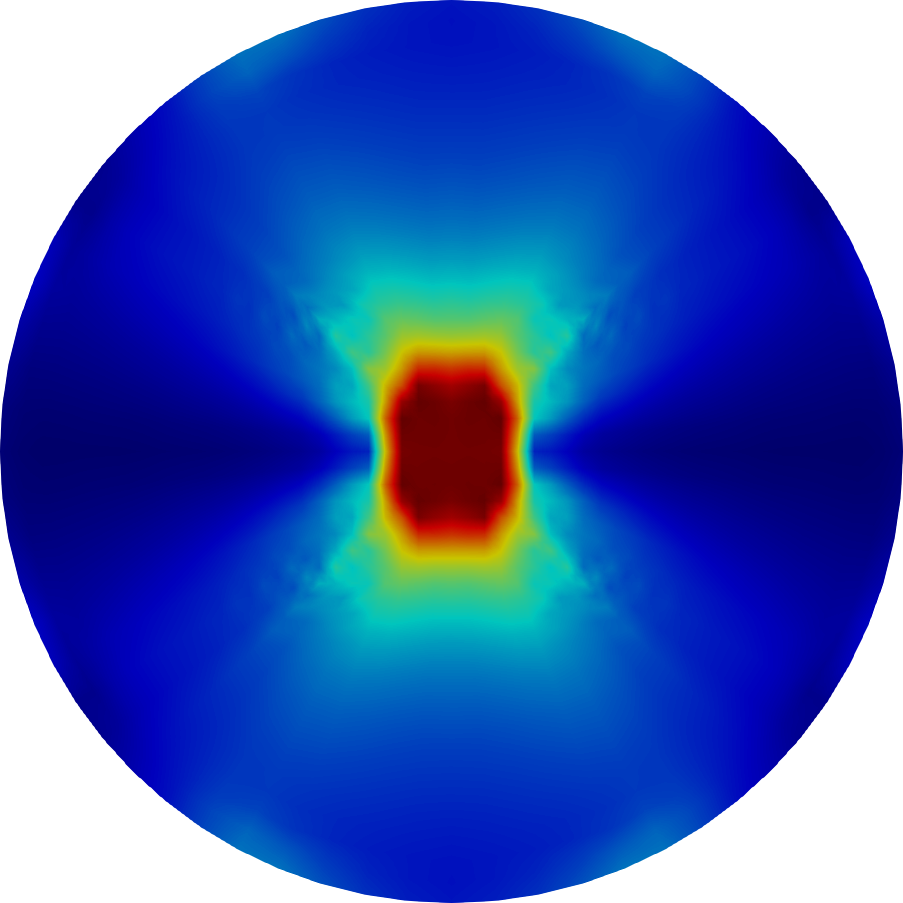}
    \caption{\small $\alpha = 1$}
  \end{subfigure}
  \hspace{0.01\textwidth}
  \begin{subfigure}[t]{0.22\textwidth}
    \includegraphics[width = \textwidth]{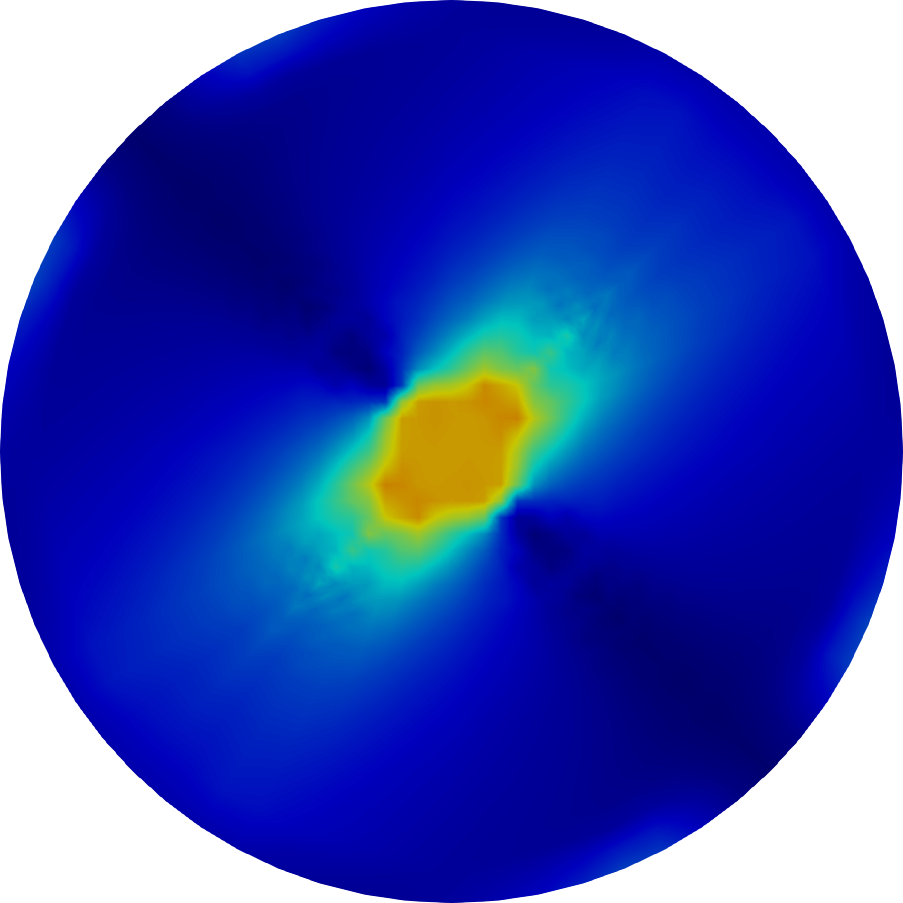}
    \caption{\small $\alpha = 2$}
  \end{subfigure}
  \hspace{0.01\textwidth}
  \begin{subfigure}[t]{0.22\textwidth}
    \includegraphics[width = \textwidth]{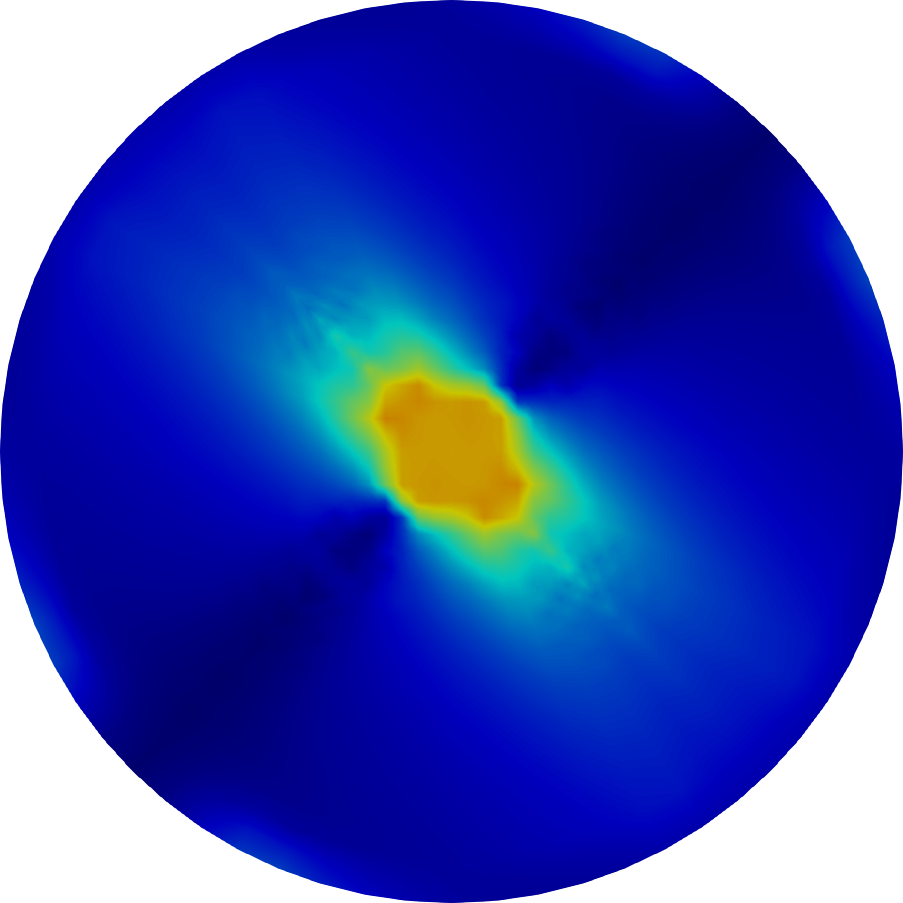}
    \caption{\small $\alpha = 3$}
  \end{subfigure}
  \hspace{0.01\textwidth}
  \begin{subfigure}[t]{0.22\textwidth}
    \includegraphics[width = \textwidth]{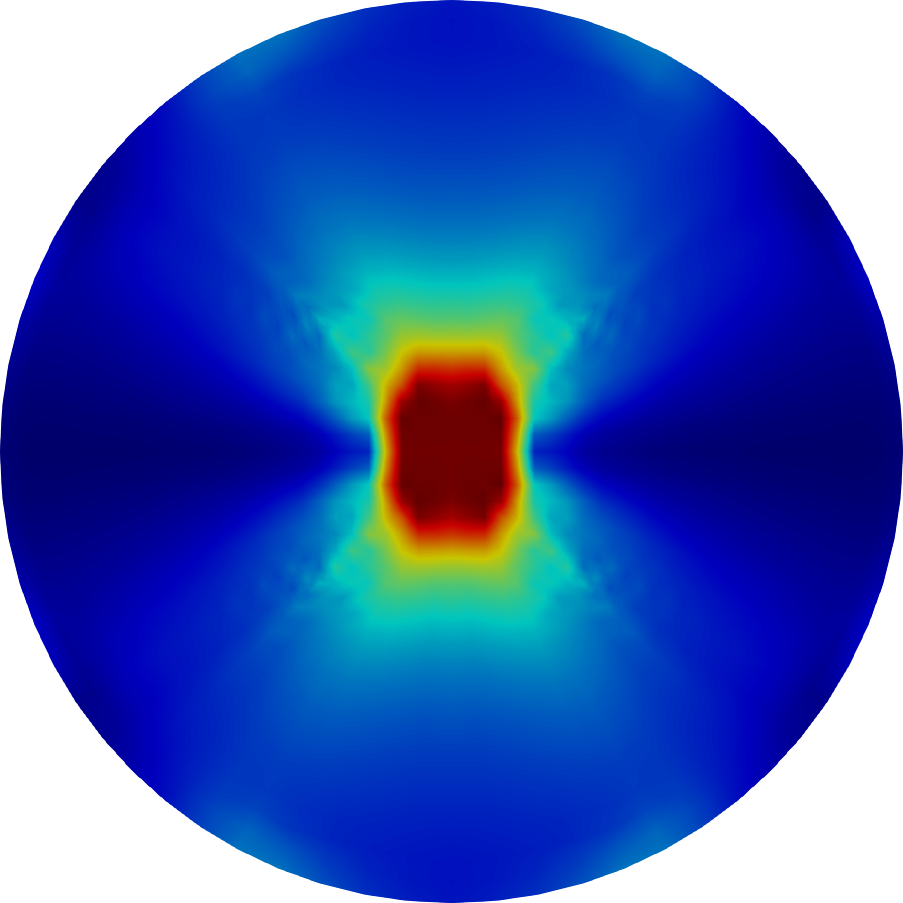}
    \caption{\small $\alpha = 4$}
  \end{subfigure}
  \\
  \vspace{5mm}
  \begin{subfigure}[t]{0.22\textwidth}
    \includegraphics[width = \textwidth]{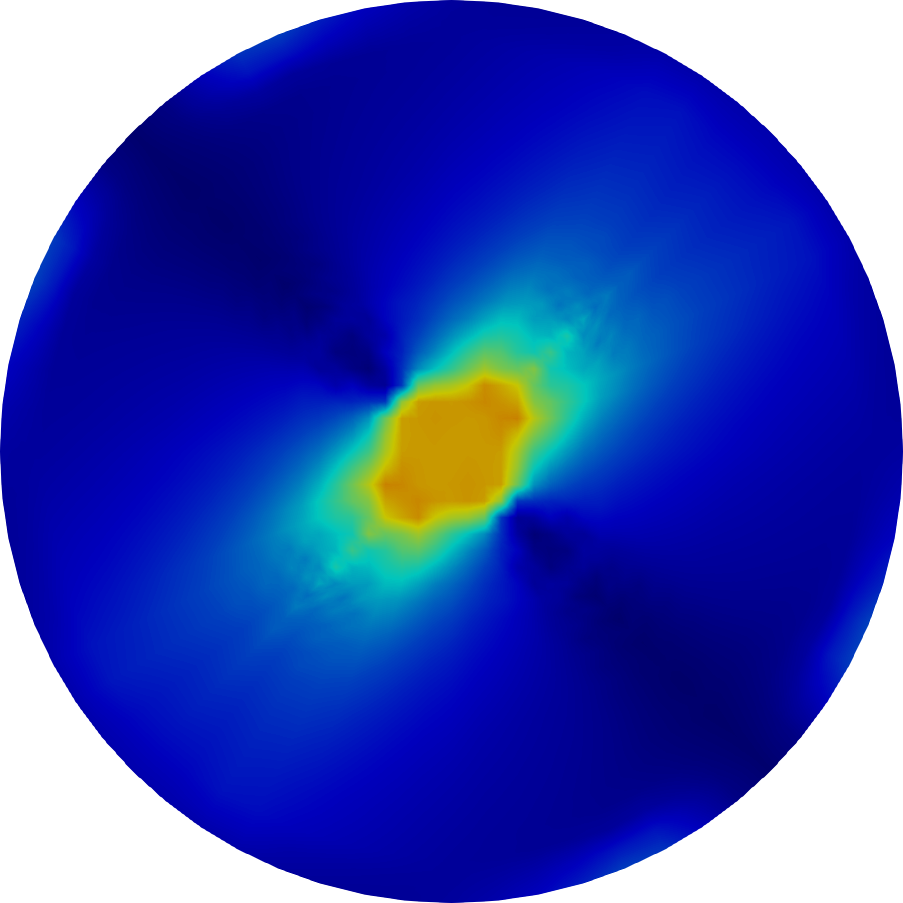}
    \caption{\small $\alpha = 5$}
  \end{subfigure}
  \hspace{0.01\textwidth}
  \begin{subfigure}[t]{0.22\textwidth}
    \includegraphics[width = \textwidth]{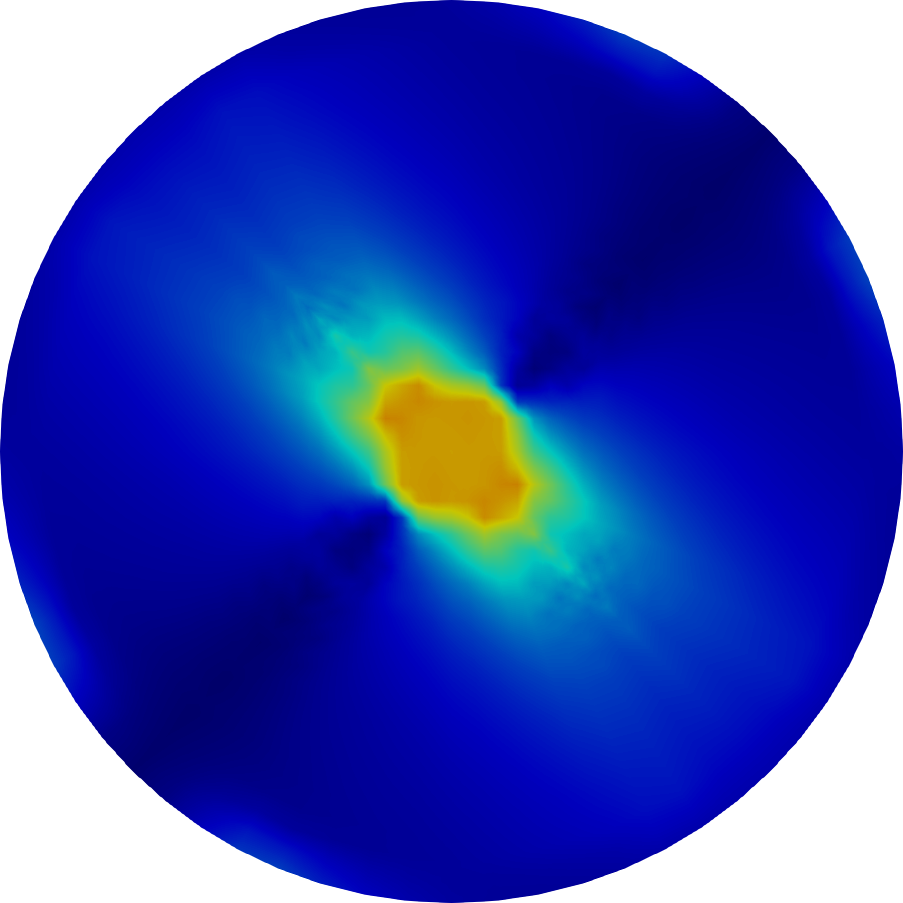}
    \caption{\small $\alpha = 6$}
  \end{subfigure}
  \hspace{0.01\textwidth}
  \begin{subfigure}[t]{0.22\textwidth}
    \includegraphics[width = \textwidth]{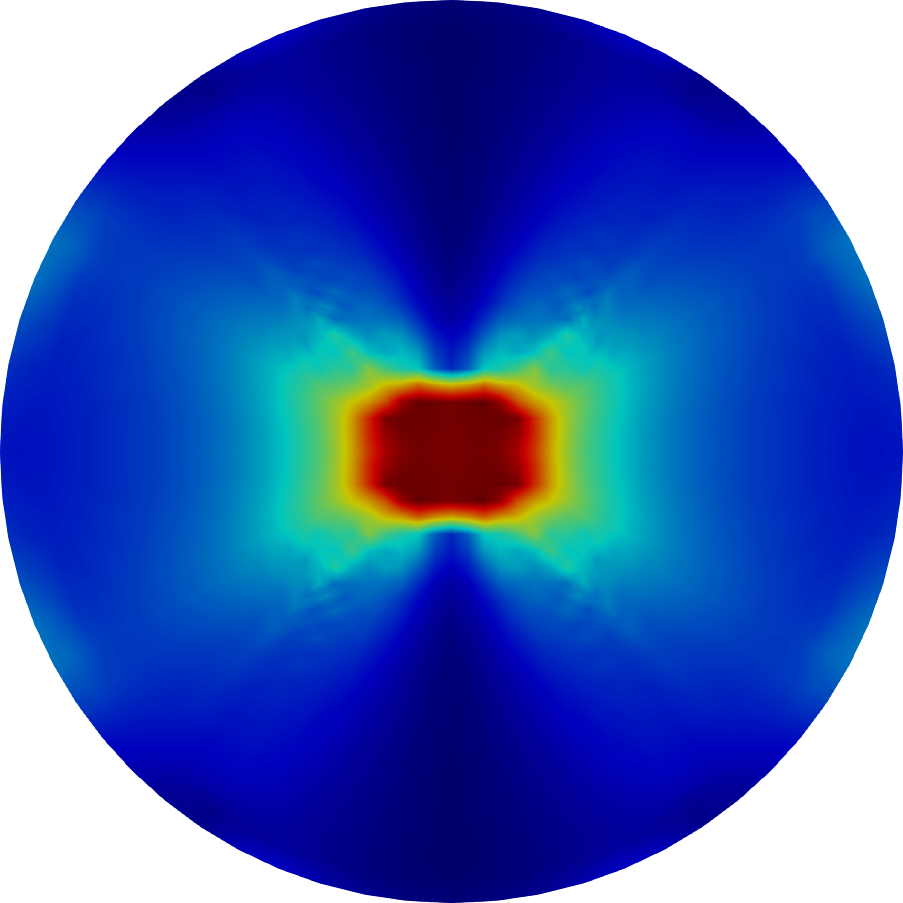}
    \caption{\small $\alpha = 7$}
  \end{subfigure}
  \hspace{0.01\textwidth}
  \begin{subfigure}[t]{0.22\textwidth}
    \includegraphics[width = \textwidth]{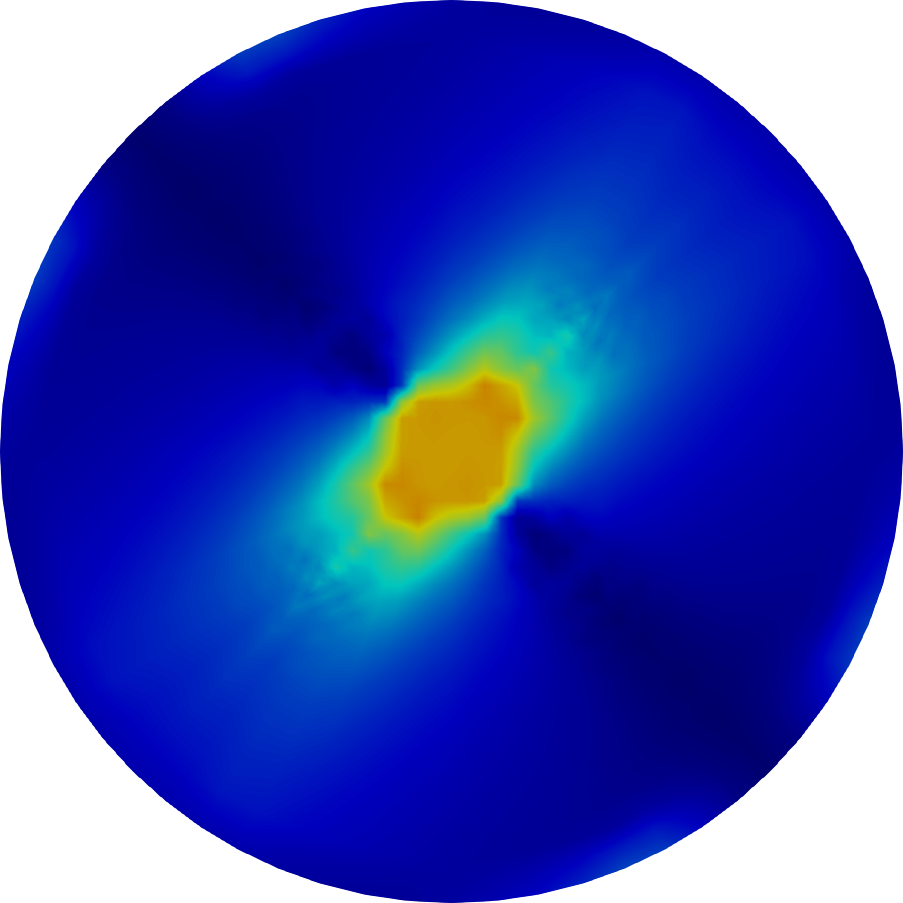}
    \caption{\small $\alpha = 8$}
  \end{subfigure}
  \\
  \vspace{5mm}
  \begin{subfigure}[t]{0.22\textwidth}
    \includegraphics[width = \textwidth]{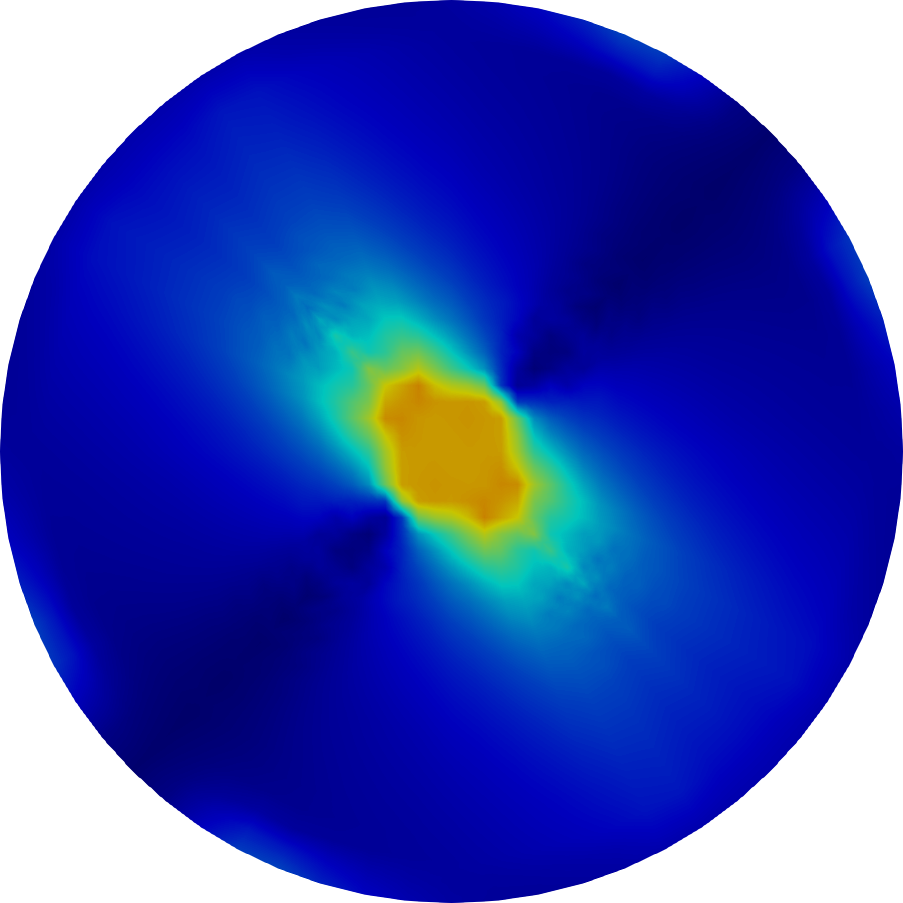}
    \caption{\small $\alpha = 9$}
  \end{subfigure}
  \hspace{0.01\textwidth}
  \begin{subfigure}[t]{0.22\textwidth}
    \includegraphics[width = \textwidth]{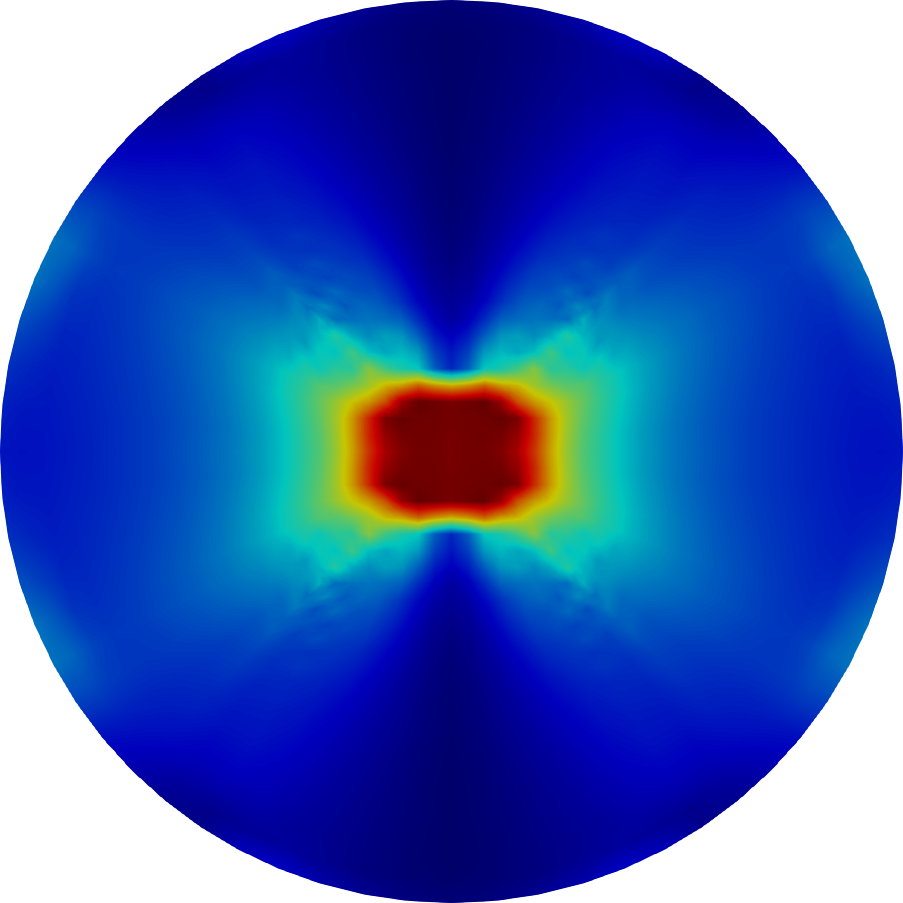}
    \caption{\small $\alpha = 10$}
  \end{subfigure}
  \hspace{0.01\textwidth}
  \begin{subfigure}[t]{0.22\textwidth}
    \includegraphics[width = \textwidth]{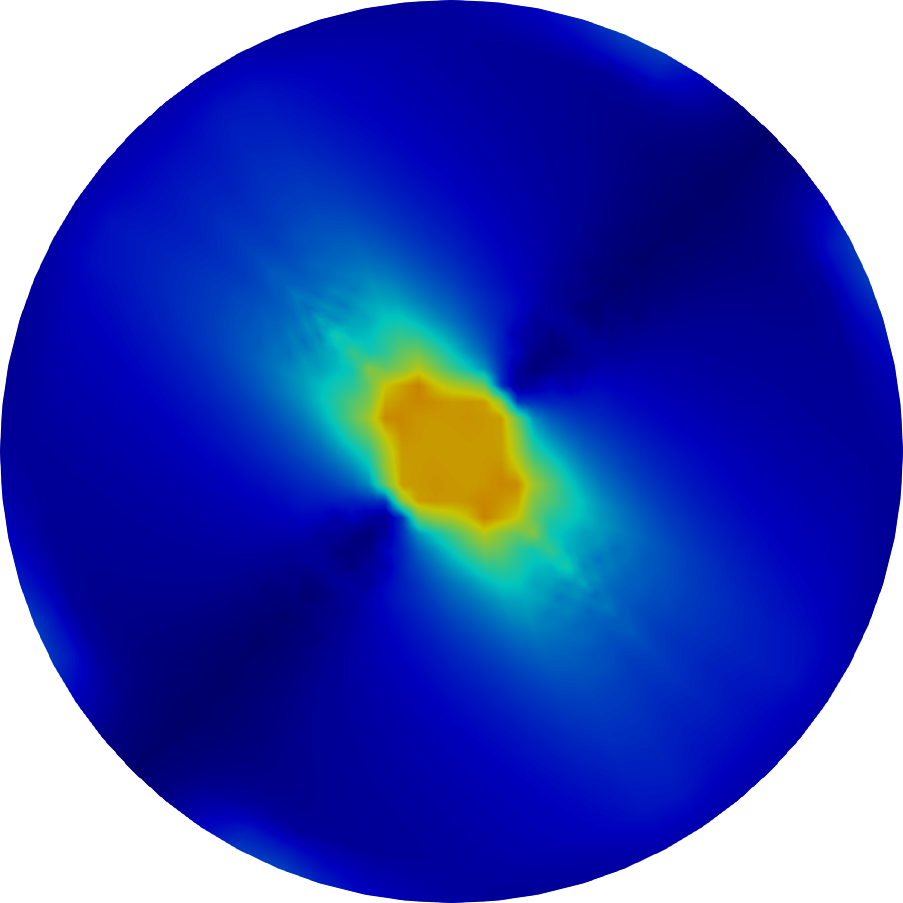}
    \caption{\small $\alpha = 11$}
  \end{subfigure}
  \hspace{0.01\textwidth}
  \begin{subfigure}[t]{0.22\textwidth}
    \includegraphics[width = \textwidth]{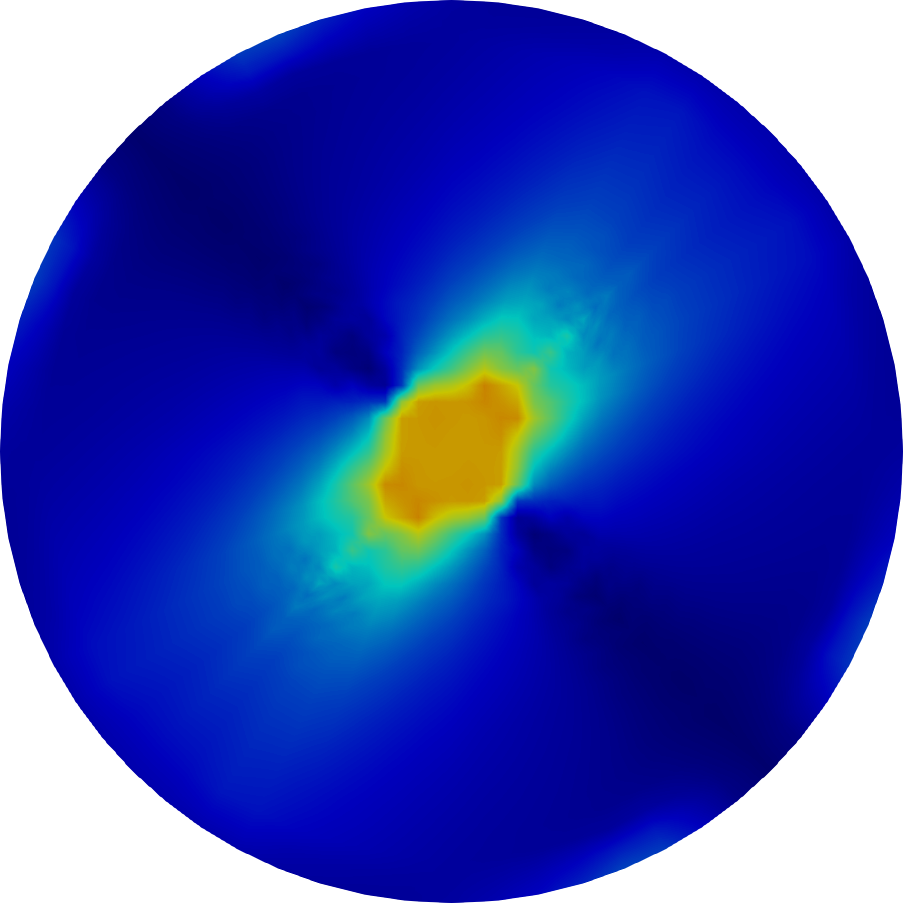}
    \caption{\small $\alpha = 12$}
  \end{subfigure}
  \\
  \vspace{5mm}
 \caption{$g^{\alpha}$ in \si{\mega\pascal} at twist of {5}{\rm mm}$^{-1}$ for $q=1.4$ and $\ell = 0.05$.}
 \label{fig:ga_l005r14}
\end{figure}

In Figures \ref{fig:ga_l005r0} and \ref{fig:ga_l005r14} the spatial distribution of $g^\alpha$ is shown for different values of $\ell$ and $q$. The distribution is very similar across all slip systems, but the absolute value increases with an increase in the combination of $\ell$ and $q$.

\begin{table}
\centering
\begin{tabular}{| c | c | c | c |}
\hline
$\ell$	&	$q$ & {\small Limiting Torque}  &
\nicefrac{$T_l(l,q)$}{$T_l(0,0)$} - 1  (\%) \\
 &  & $T_l (\si{\nano\newton\meter})$ & \\
\hline
0&0& 202863 &-\\
0.05&0& 209394&3.22\\ 
0.10&0& 215967&6.46\\ 
0.05&1& 210812&3.92\\ 
0.10&1& 218920&7.92\\ 
0.05&1.4& 211383&4.2\\ 
0.10&1.4& 220129&8.51\\ 
\hline
\end{tabular}
\caption{Limiting torque $T_l = T$ at $\kappa = 5 mm^{-1}$, relative to the torque at the initialization of yielding, for different values of the length scale $\ell$ and self/latent ratio $q$}
\label{tab:torque}
\end{table}


\section{Concluding remarks}

This work has been concerned with the formulation of a thermodynamically consistent theory of single-crystal stress-gradient plasticity, and exploration of its behaviour by application to the problem of simple torsion. A viscoplastic approximation of the model has been implemented, using finite elements. Results of dimensionless torque against twist show a clear size-dependent strengthening, in the form of an increase in incipient yield and subsequent `lifting' of the torque-twist curves. Minimal hardening is evident.

There are a number of further exploratory studies that would be add to an understanding the model presented here. One such class of problems would be in the domain of polycrystalline plasticity. For these and other examples, the nature of the size-dependent response - strengthening and/or hardening would be of interest.

A further area of interest would be in relation to alternative forms of the model of stress-gradient plasticity: for example, models even closer to that in \cite{chakravarthy-curtin2011}, in the sense of \eqref{CCyield}. An associated challenge would be to do so in a thermodynamically consistent way.

\section{Acknowledgements}
BDR acknowledges support for this work from the National Research Foundation, through the South African Research Chair in Computational Mechanics. PS acknowledges support by the Royal Society, UK, through a Wolfson Research Merit Award and the German Science Foundation through the project P10 of the Priority Programme 2013 on Residual Stresses as well as together with AK through the project C5 of the Collaborative Research Centre 814 on Additive Manufacturing.


\end{document}